\documentclass[aps,prl,preprintnumbers,twocolumn,groupedaddress,nofootinbib]{revtex4}

\usepackage[dvips]{graphicx}
\usepackage{color}
\usepackage{amsmath,amssymb,slashed, amsthm}
\usepackage{hyperref}

\usepackage{changepage}

\usepackage{tikz}
\usetikzlibrary{calc} 
\usetikzlibrary{patterns,snakes} 
\usetikzlibrary{decorations.pathreplacing} 
\usetikzlibrary{decorations.markings} 
\usetikzlibrary{decorations.pathmorphing} 
\usetikzlibrary{positioning}
\usetikzlibrary{arrows.meta}

\newif\ifdraft
\drafttrue
\newif\ifpreprint
\preprinttrue
\def\spa#1.#2{\left\langle#1\,#2\right\rangle}
\def\spb#1.#2{\left[#1\,#2\right]}

\font\tenshuffle=shuffle10 \font\sevenshuffle=shuffle7 \font\fiveshuffle=shuffle7 at 5pt
\def\shuffle{{%
\def\Dshuffle{\mathbin{\hbox{\tenshuffle\char'001}}}%
\def\Sshuffle{\mathbin{\hbox{\sevenshuffle\char'001}}}%
\def\SSshuffle{\mathbin{\hbox{\fiveshuffle\char'001}}}%
\mathchoice{\Dshuffle}{\Dshuffle}{\Sshuffle}{\SSshuffle}}}

\def\beq{\begin{equation}}
\def\eeq{\end{equation}}

\let\Im\relax

\DeclareMathOperator{\Im}{Im}

\newcommand{\eq}{\begin{equation}}
\newcommand{\eqe}{\end{equation}}
\newcommand{\eqa}{\begin{eqnarray}}
\newcommand{\eqae}{\end{eqnarray}}

\newcommand{\p}{\partial}

\newcommand{\ep}{\epsilon}

\newcommand{\bea}{\begin{eqnarray}}
\newcommand{\eea}{\end{eqnarray}}

\newcommand{\bma}{\begin{matrix}}
\newcommand{\ema}{\cr\end{matrix}}

\newcommand{\ZZ}{\mathbb Z}

\def\mA{\mathfrak{A}}
\def\mB{\mathfrak{B}}

\def\cA{{\cal A}}

\def\cG{{\cal G}}

\def\cK{{\cal K}}

\def\mA{\mathfrak{A}}
\def\mB{\mathfrak{B}}
\def\mC{\mathfrak{C}}

\def\mJ{\mathfrak{J}}

\def\ZZ{{\mathbb Z}}

\def\Im{{\rm Im \,}}

\def\half{{1\over 2}}
\def\thalf{{\tfrac{1}{2}}}
\def\p{\partial}

\def\tet{\vartheta}
\def\ep{\varepsilon}
\def\om{\omega}

\def\sm{\smallskip}
\def\no{\nonumber}

\def\qq{q}
\def\Del{\Delta}

\def\mylength{-0.35cm}
\def\myotherlength{-0.25cm}

\newbox\charbox
\newbox\slabox
\def\s#1{{     % Feynman slash
        \setbox\charbox=\hbox{$#1$}
        \setbox\slabox=\hbox{$/$}
        \dimen\charbox=\ht\slabox
        \advance\dimen\charbox by -\dp\slabox
        \advance\dimen\charbox by -\ht\charbox
        \advance\dimen\charbox by \dp\charbox
        \divide\dimen\charbox by 2
        \raise-\dimen\charbox\hbox to \wd\charbox{\hss/\hss}
        \llap{$#1$}
}}

\makeatletter
\renewcommand{\p@subsection}{}
\renewcommand{\p@subsubsection}{}
\makeatother

\begin{document}

\preprint{UUITP--07/25}

\title{Meromorphic higher-genus integration kernels via convolution over homology cycles}

\author{Eric D'Hoker$^a$, and
Oliver Schlotterer$^{b}$\footnote{Corresponding author, oliver.schlotterer@physics.uu.se}}
\affiliation{$^a$ Mani L.\ Bhaumik Institute for Theoretical Physics,  Department of Physics and Astronomy,  University of California, Los Angeles,  CA 90095, USA}
\affiliation{$^b$ Department of Physics and Astronomy, 
Department of Mathematics,
Centre for Geometry and Physics,
Uppsala University, Box 516, 75120 Uppsala, Sweden}

\begin{abstract}
Polylogarithms on arbitrary higher-genus Riemann surfaces can be constructed from meromorphic integration kernels with at most simple poles, whose definition was given by Enriquez via functional properties. In this work, homotopy-invariant convolution integrals over homology cycles are shown to provide a direct construction of Enriquez kernels solely from holomorphic Abelian differentials and the prime form. Our new representation is used to demonstrate the closure of the space of Enriquez kernels under convolution over homology cycles and under variations of the moduli.
The results of this work further strengthen the remarkable parallels of Enriquez kernels with the non-holomorphic modular tensors recently developed in an alternative construction of higher-genus polylogarithms.
\end{abstract}

\maketitle

%%%%%%%%%%%%%%%%%%%%%%%%%%%%%%%%%%%%%%%%%%%%%%%%%%%%%%%%%
%%%%%%%%%%%%%%%%%%%%%%%%%%%%%%%%%%%%%%%%%%%%%%%%%%%%%%%%%
\section{Introduction}
%%%%%%%%%%%%%%%%%%%%%%%%%%%%%%%%%%%%%%%%%%%%%%%%%%%%%%%%%
%%%%%%%%%%%%%%%%%%%%%%%%%%%%%%%%%%%%%%%%%%%%%%%%%%%%%%%%%
\vspace{\myotherlength}

Polylogarithms and their elliptic counterparts provide a unifying mathematical framework that has considerably extended 
our computational reach in quantum field theory and string theory in the recent past \cite{Bourjaily:2022bwx, Berkovits:2022ivl, Abreu:2022mfk}. While these types of polylogarithms suffice to organize iterated integrals on the sphere and the torus, problems at the cutting edge of collider physics,  gravitational wave physics, and string theory call for polylogarithms on varieties of more general topology.

\sm

Several formulations of polylogarithms on a Riemann surface of arbitrary genus have been introduced in terms of different integration kernels, which may be combined into flat connections, and generate function spaces that close under integration \cite{Enriquez:2011, Enriquez:2021, Ichikawa:2022qfx, Enriquez:2022, DHoker:2023vax, Baune:2024, DHoker:2025szl, Enriquez:next}. 
Disposing of several different spaces of integration kernels is  familiar from the elliptic case  where integration kernels can be multiple-valued and meromorphic \cite{Levin:1997, Levin:2007, CEE, Broedel:2017kkb}, single-valued and non-meromorphic \cite{BrownLevin, Broedel:2014vla}, or single-valued and meromorphic allowing for double poles  \cite{Enriquez:2023}.

\sm

The protagonists of this work are the meromorphic integration kernels for a Riemann surface  of arbitrary genus that were defined through their functional properties by Enriquez in \cite{Enriquez:2011}. A method for their calculation was given in \cite{Enriquez:2021} where also low rank examples were worked out; a representation in terms of Poincar\'e series was given for hyper-elliptic surfaces in \cite{Baune:2024}; and their relation with non-meromorphic modular tensors was exhibited in \cite{DHoker:2025szl}. 

\sm

In this paper, we shall provide a simple direct recursion relation for Enriquez kernels in terms of convolution integrals over homology $\mA$ cycles of the surface $\Sigma$. Their integrands are built solely from holomorphic Abelian differentials and  combinations of the prime form that may be expressed in terms of Riemann theta functions only. This representation of Enriquez kernels is used to show closure of  their function space under convolution on homology $\mA$ cycles;  to evaluate their variation with respect to the moduli of $\Sigma$;  and to exhibit their close parallels with the single-valued integration kernels of~\cite{DHoker:2023vax}.

\sm

Our representation of Enriquez kernels is expected to facilitate evaluating them numerically;
calculating their behavior under degenerations of $\Sigma$; and generalizing the representation of elliptic polylogarithms as iterated integrals of modular forms \cite{Broedel:2015hia, Adams:2017ejb, Broedel:2018iwv} 
beyond genus one. These features should play a key role for applications of higher-genus polylogarithms to Feynman integrals \cite{Huang:2013kh, Georgoudis:2015hca, Doran:2023yzu, Marzucca:2023gto, Jockers:2024tpc, Duhr:2024uid} and to string amplitudes in their chiral-splitting formulation~\cite{DHoker:1988pdl, DHoker:2020prr, DHoker:2021kks, DHoker:2023khh}. Moreover, the construction of meromorphic integration kernels from convolution integrals over homology cycles should be of use for the integrals on higher-dimensional varieties encountered in recent precision calculations for particle physics \cite{Bourjaily:2022bwx, Forner:2024ojj} and gravity \cite{Frellesvig:2023bbf, Klemm:2024wtd, Driesse:2024feo, Frellesvig:2024rea}.

\vspace{\mylength}
%%%%%%%%%%%%%%%%%%%%%%%%%%%%%%%%%%%%%%%%%%%%%%%%%%%%%%%%%
%%%%%%%%%%%%%%%%%%%%%%%%%%%%%%%%%%%%%%%%%%%%%%%%%%%%%%%%
\section{Enriquez kernels}
\label{sec:2}
%%%%%%%%%%%%%%%%%%%%%%%%%%%%%%%%%%%%%%%%%%%%%%%%%%%%%%%%%
%%%%%%%%%%%%%%%%%%%%%%%%%%%%%%%%%%%%%%%%%%%%%%%%%%%%%%%%%
\vspace{\myotherlength}

\begin{figure}[htb]
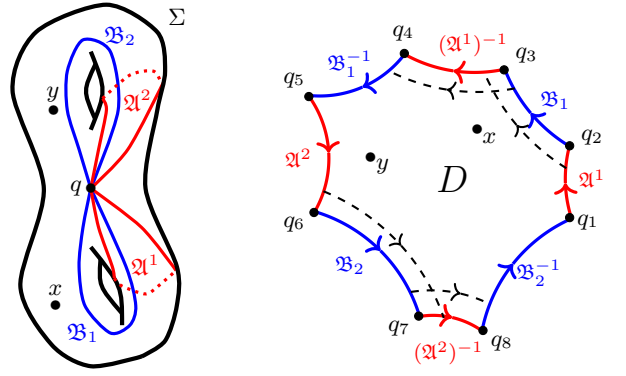

\begin{center}
\tikzpicture[scale=0.65]
\scope[xshift=2.2cm,yshift=0cm, scale=0.8, rotate=90]
% handles
\draw[ultra thick] (1.5,1.2) .. controls (2.5, 1.2) .. (3.5,2);
\draw[ultra thick] (1.8,1.2) .. controls (2.5, 1.8) .. (3.2,1.75);
\draw[ultra thick] (6.5,2) .. controls (7.5, 1.6) .. (8.5,2);
\draw[ultra thick] (6.8,1.9) .. controls (7.5, 2.2) .. (8.2,1.87);
% cycles
\draw[very thick, color=blue] plot [smooth] coordinates {(5,2) (4,2.2) (3, 2.3) (2, 2.15) (1.3,1.5) (1.5,1) ( 2, 0.85) (3,1.1) (5,2) };
\draw[very thick, color=blue] plot [smooth] coordinates {(5,2) (6,2.3) (7, 2.6) (8, 2.6) (8.8,2.2) (8.5,1.5)  (7,1.3) (5,2) };
\draw[very thick, color=red] plot [smooth] coordinates { (5,2) (4,1.8) (3,1.45) (2.7,1.4)};
\draw[very thick, color=red, dotted] plot [smooth] coordinates { (2.7,1.4) (2.5,1.2) (2.3,0.6) (2.5, 0.275) (2.9, -0.18)};
\draw[very thick, color=red] plot [smooth] coordinates { (2.9,-0.18) (4,0.7) (5,2)};
\draw[very thick, color=red] plot [smooth] coordinates { (5,2) (6,1.8) (7,1.6) (7.3,1.7)};
\draw[very thick, color=red, dotted] plot [smooth] coordinates { (7.3,1.7) (7.9,1) (8,0.4) (7.7, 0.2)};
\draw[very thick, color=red] plot [smooth] coordinates { (7.7, 0.15) (6,1.05) (5,2)};
% contour
\draw[ultra thick] plot [smooth] coordinates 
{(0.7,0.7) (0.45,2) (1,3) (2.5, 3.8)  (5,3.2) (7.5, 4) (9,3.4) (9.7,1.6) 
(9,0.3) (8,0.1) (7, 0.3) (5,0.6) (3, -0.2) (2,-0.2) (1.1,0.1) (0.7,0.7) };
% labels
\draw [color=red] (3.05,0.6) node{\small $\mA^1$};
\draw [color=red] (7.25,0.8) node{\small $\mA^2$};
\draw [color=blue] (1.3,2.2) node{\small $\mB_1$};
\draw [color=blue]  (8.9,1.25) node{\small $\mB_2$};
\draw (9.4,-0.2) node{{$\Sigma$}};
% base point 
\draw (5,2) node{$\bullet$};
\draw (5,2.38) node{ $\qq$};
\draw (2,2.9) node{ $\bullet$};
\draw (2.4,2.9) node{ $x$};
\draw (7,2.95) node{ $\bullet$};
\draw (7.4,2.97) node{ $y$};
\endscope
\scope[xshift=8cm,yshift=4.2cm, scale=4.2]
%\draw  [thick, domain=0:90] plot ({cos(\x)}, {sin(\x)});
%\draw  [thick, domain=90:180] plot ({cos(\x)}, {sin(\x)});
%\draw  [thick, domain=180:270] plot ({cos(\x)}, {sin(\x)});
%\draw  [thick, domain=270:360] plot ({cos(\x)}, {sin(\x)});
\draw  [very  thick, color=red, <-, domain=182:198] plot ({cosh(0.6)*cos(0)+sinh(0.6)*cos(\x)}, {cosh(0.6)*sin(0)+sinh(0.6)*sin(\x)});
\draw  [very  thick, color=red, domain=166:182] plot ({cosh(0.6)*cos(0)+sinh(0.6)*cos(\x)}, {cosh(0.6)*sin(0)+sinh(0.6)*sin(\x)});
\draw  [very  thick, color=red, <-, domain=261.5:282] plot ({cosh(0.65)*cos(85)+sinh(0.65)*cos(\x)}, {cosh(0.65)*sin(85)+sinh(0.65)*sin(\x)});
\draw  [very  thick, color=red,  domain=241:261.5] plot ({cosh(0.65)*cos(85)+sinh(0.65)*cos(\x)}, {cosh(0.65)*sin(85)+sinh(0.65)*sin(\x)});
\draw  [very  thick, color=red, <-, domain=363:396] plot ({cosh(0.5)*cos(175)+sinh(0.5)*cos(\x)}, {cosh(0.5)*sin(175)+sinh(0.5)*sin(\x)});
\draw  [very  thick, color=red,  domain=330:363] plot ({cosh(0.5)*cos(175)+sinh(0.5)*cos(\x)}, {cosh(0.5)*sin(175)+sinh(0.5)*sin(\x)});
\draw  [very  thick, color=red, <-, domain=77:100] plot ({cosh(0.4)*cos(265)+sinh(0.4)*cos(\x)}, {cosh(0.4)*sin(265)+sinh(0.4)*sin(\x)});
\draw  [very  thick, color=red,  domain=54:77] plot ({cosh(0.4)*cos(265)+sinh(0.4)*cos(\x)}, {cosh(0.4)*sin(265)+sinh(0.4)*sin(\x)});
\draw  [very  thick, color=blue, <-, domain=220:239] plot ({cosh(0.7)*cos(40)+sinh(0.7)*cos(\x)}, {cosh(0.7)*sin(40)+sinh(0.7)*sin(\x)});
\draw  [very  thick, color=blue, domain=201:220] plot ({cosh(0.7)*cos(40)+sinh(0.7)*cos(\x)}, {cosh(0.7)*sin(40)+sinh(0.7)*sin(\x)});
\draw  [very  thick, color=blue, <-, domain=293.5:323] plot ({cosh(0.5)*cos(125)+sinh(0.5)*cos(\x)}, {cosh(0.5)*sin(125)+sinh(0.5)*sin(\x)});
\draw  [very  thick, color=blue, domain=264:293.5] plot ({cosh(0.5)*cos(125)+sinh(0.5)*cos(\x)}, {cosh(0.5)*sin(125)+sinh(0.5)*sin(\x)});
\draw  [very  thick, color=blue, <-, domain=45:73] plot ({cosh(0.7)*cos(225)+sinh(0.7)*cos(\x)}, {cosh(0.7)*sin(225)+sinh(0.7)*sin(\x)});
\draw  [very  thick, color=blue, domain=17:45] plot ({cosh(0.7)*cos(225)+sinh(0.7)*cos(\x)}, {cosh(0.7)*sin(225)+sinh(0.7)*sin(\x)});
\draw  [very  thick, color=blue, <-, domain=141.5:169] plot ({cosh(0.7)*cos(315)+sinh(0.7)*cos(\x)}, {cosh(0.7)*sin(315)+sinh(0.7)*sin(\x)});
\draw  [very  thick, color=blue, domain=114:141.5] plot ({cosh(0.7)*cos(315)+sinh(0.7)*cos(\x)}, {cosh(0.7)*sin(315)+sinh(0.7)*sin(\x)});
% identifications
\draw  [thick, color=black, dashed, ->, domain=238:220] plot ({cosh(0.9)*cos(40)+sinh(0.9)*cos(\x)}, {cosh(0.9)*sin(40)+sinh(0.9)*sin(\x)});
\draw  [thick, color=black, dashed,  domain=220:203] plot ({cosh(0.9)*cos(40)+sinh(0.9)*cos(\x)}, {cosh(0.9)*sin(40)+sinh(0.9)*sin(\x)});
\draw (0.148,-0.74) [fill=black] circle(0.02cm) ;
\draw  [thick, color=black, dashed, ->, domain=68:42] plot ({cosh(0.9)*cos(225)+sinh(0.9)*cos(\x)}, {cosh(0.9)*sin(225)+sinh(0.9)*sin(\x)});
\draw  [thick, color=black, dashed,  domain=42:19] plot ({cosh(0.9)*cos(225)+sinh(0.9)*cos(\x)}, {cosh(0.9)*sin(225)+sinh(0.9)*sin(\x)});
\draw  [thick, color=black, dashed, ->, domain=281:261] plot ({cosh(0.85)*cos(85)+sinh(0.85)*cos(\x)}, {cosh(0.85)*sin(85)+sinh(0.85)*sin(\x)});
\draw  [thick, color=black, dashed, domain=261:243] plot ({cosh(0.85)*cos(85)+sinh(0.85)*cos(\x)}, {cosh(0.85)*sin(85)+sinh(0.85)*sin(\x)});
\draw  [thick, color=black, dashed, ->, domain=99:77] plot ({cosh(0.6)*cos(265)+sinh(0.6)*cos(\x)}, {cosh(0.6)*sin(265)+sinh(0.6)*sin(\x)});
\draw  [thick, color=black, dashed, domain=77:63] plot ({cosh(0.6)*cos(265)+sinh(0.6)*cos(\x)}, {cosh(0.6)*sin(265)+sinh(0.6)*sin(\x)});
\draw (0.57,0.16) [fill=black] circle(0.02cm) ;
\draw (0.251,0.53) [fill=black] circle(0.02cm) ;
\draw (-0.7,0.4) [fill=black] circle(0.02cm) ;
\draw (-0.235,0.61) [fill=black] circle(0.02cm) ;
\draw (-0.675,-0.165) [fill=black] circle(0.02cm) ;
\draw (-0.165,-0.67) [fill=black] circle(0.02cm) ;
\draw (0.57,-0.19) [fill=black] circle(0.02cm) ;

\draw (0.66, -0.21) node{\small $\qq_1$};
\draw (0.68, 0.2) node{\small $\qq_2$};
\draw (0.36, 0.6) node{\small $\qq_3$};
\draw (-0.24, 0.71) node{\small $\qq_4$};
\draw (-0.77, 0.48) node{\small $\qq_5$};
\draw (-0.77, -0.22) node{\small $\qq_6$};
\draw (-0.26, -0.73) node{\small $\qq_7$};
\draw (0.24, -0.8) node{\small $\qq_8$};

\draw [color=red] (0.68,0) node{\footnotesize $\mA^1$};
\draw [color=blue] (0.5,0.38) node{\footnotesize $\mB_1$};
\draw [color=red] (0.1,0.65) node{\footnotesize $(\mA^1)^{-1}$};
\draw [color=blue] (-0.5,0.56) node{\footnotesize $\mB_1^{-1}$};
\draw [color=red] (-0.75,0.1) node{\footnotesize $\mA^2$};
\draw [color=blue]  (-0.5,-0.42) node{\footnotesize $\mB_2$};
\draw [color=red] (-0.02,-0.85) node{\footnotesize $(\mA^2)^{-1}$};
\draw [color=blue]  (0.42,-0.46) node{\footnotesize $\mB_2^{-1}$};
\draw (0,0) node{{\Large $D$}};
\draw (0.12,0.24) node{{$\bullet$}};
\draw (0.18,0.2) node{{$x$}};
\draw (-0.4,0.1) node{{$\bullet$}};
\draw (-0.34,0.06) node{{$y$}};
\endscope
%%%
%%%

\endtikzpicture
\caption{The left panel represents  a compact genus-two Riemann surface $\Sigma$ and a choice of canonical homology cycles $\mA^1, \mA^2, \mB_1, \mB_2$ with a common base point $q$. The right panel represents a fundamental domain $D \subset \tilde \Sigma$ for the action of $ \pi_1(\Sigma, \qq)$ on $\Sigma$, which can be obtained by cutting $\Sigma$ along the cycles in the left panel. The surface $\Sigma$ may be reconstructed from $D$ by pairwise identifying inverse boundary components with one another under the dashed arrows. The projection $\pi: \tilde \Sigma \to \Sigma$ maps the vertices $\qq_1, \cdots ,q_8 \in D$ to the point~$\qq \in \Sigma$ and the points $x,y$ to their images by the same name.
 \label{fig:1}}
\end{center}
\end{figure}

We consider a compact Riemann surface $\Sigma$ of arbitrary genus $h \geq 1$, denote its universal covering space by $\tilde \Sigma$,  the associated projection by $\pi : \tilde \Sigma \to \Sigma $, and the intersection pairing of the homology group $H_1(\Sigma, \ZZ)$ by $\mJ$. A canonical homology basis of cycles $\mA^I$ and $\mB_I$ obeys $\mJ(\mA^I, \mA^J) = \mJ (\mB_I, \mB_J)=0$ and $\mJ(\mA^I, \mB_J)=\delta ^I_J$ for $I,J \in \{ 1, \cdots, h\}$, as shown in figure \ref{fig:1}. The holomorphic Abelian differentials $\om_J(x)$ for $x\in \Sigma$ are normalized on $\mA^I$ cycles and define the period matrix $\Omega$ of the surface $\Sigma$~via,
\bea
\oint _{\mA^I} \om_J = \delta ^I_J \hskip 0.7in \oint _{\mB_I} \om_J = \Omega _{IJ}
\label{permat}
\eea
Choosing the cycles $\mA^I$ and $\mB_J$ so that they share a~common base point $ q \in \Sigma$ as in figure \ref{fig:1} promotes them into a set of generators of the first homotopy group $\pi_1(\Sigma, q)$ of $\Sigma$.

\sm

In \cite{Enriquez:2011}, Enriquez introduced a meromorphic connection $d - \cK_\text{E}$ on $\tilde \Sigma$ with simple poles, that is valued in a freely generated Lie algebra out of which meromorphic polylogarithms may be systematically constructed \cite{Baune:2024, DHoker:2024ozn, Enriquez:next}.  Here we shall be principally interested in the coefficient functions $g^{I_1 \cdots I_r}{}_J(x,y)$, for $x,y\in \tilde \Sigma, \ r \geq 0$ and $I_1, \cdots, I_r \in \{ 1, \cdots, h\}$ that arise as the connection form $\cK_\text{E}$ is Taylor expanded in the $2h$ generators $a^1,\cdots,a^h$ and $b_1,\cdots,b_h$ of the Lie algebra,
\bea
\cK_\text{E}(x,y) =  \sum_{r=0}^\infty  g^{I_1  \cdots I_r}{}_J(x,y)  
[ b_{I_1},[b_{I_2},{\cdots} [ b_{I_r},a^J   ] {\cdots} ]]
\label{keform}
\eea
 and are related to those used in \cite{Enriquez:2011} by $g^{I_1\cdots I_r}{}_J(x,y) = (-2\pi i)^r \omega^{I_1\cdots I_r}{}_J(x,y)$. Throughout, we shall adopt the Einstein summation convention under which a pair of repeated upper and lower indices is understood to be summed without exhibiting the summation symbol, such as $J,I_1,\cdots,I_r$ in (\ref{keform}).
 The coefficient functions $g^{I_1\cdots I_r}{}_J(x,y) $ in (\ref{keform}), which are also referred to as Enriquez kernels, are uniquely defined by the following properties \cite{Enriquez:2011, DHoker:2025szl}:

\begin{enumerate}
\itemsep=0in
\item The Enriquez kernel $g^{I_1 \cdots I_r}{}_J(x,y)$ is a $(1,0)$ form in  $x \in \tilde \Sigma$ and a scalar in $y \in \tilde \Sigma$ which is meromorphic for $x,y \in \tilde \Sigma$ and locally holomorphic in the complex moduli of $\Sigma$ (whose dependence will be suppressed throughout).
\item In the preferred fundamental domain $D$, see the right panel of figure \ref{fig:1}, $g^{I_1 \cdots I_r}{}_J(x,y)$ is holomorphic in $x,y \in D$ for $r \geq 2$,  has a simple pole in $x$ at $y$ only at $r=1$,
\beq
g^I{}_J(x,y) = \frac{\delta^I_J \, dx}{x-y} + {\rm reg}
\label{polegij}
\eeq
 and is given by $g^\emptyset {}_J(x,y) = \om_J(x)$ for $r=0$. 
 \item The monodromies in $x$ of $g^{I_1 \cdots I_r}{}_J(x,y)$  around $\mA$ cycles are trivial, and around the cycle $\mB_L$ are given by, 
 \begin{align}
\label{2.mongx}
&g^{I_1 \cdots I_r}{}_J(\mB_L  \cdot  x,y)  = 
g^{I_1 \cdots I_r}{}_J(x,y)   \\
&\quad + \sum_{k=1}^r { (-2\pi i)^k \over k!} \, \delta ^{I_1 \cdots I_k} _L \, g^{I_{k+1} \cdots I_r}{}_J(x,y)
\notag
\end{align}
where $\mB_L \cdot x$ denotes the action of the element $\mB_L \in \pi _1(\Sigma , q)$ on the point $x \in \Sigma$ and  the generalized Kronecker symbol is defined by $\delta^{I_{1}  \cdots I_k}_L=\delta^{I_{1}}_L  \cdots \delta^{ I_k}_L$. 
\end{enumerate}
The forms $g^{I_1 \cdots I_r}{}_J(x,y)$ may have poles in $x$ at $\pi^{-1}(y)$ for all $r \geq 1$, as mandated by the monodromy relations.  Important consequences derived in \cite{Enriquez:2011,DHoker:2025szl} are as follows.

\begin{enumerate}
\itemsep=0in
\item trivial $\mA$ monodromies in the variable $y$, and $\mB$ monodromies given by, 
\begin{align}
\label{2.mongy}
&g^{I_1 \cdots I_r}{}_J(x ,\mB_L \cdot y)  =   g^{I_1 \cdots I_r}{}_J(x,y)  \\
&\quad
+\delta ^{I_r}_J \sum_{k=1}^r { (2\pi i)^k \over k !} \, g^{I_1 \cdots I_{r-k}} {}_L(x,y) \, \delta^{I_{r-k+1} \cdots I_{r-1}}_L 
\notag
\end{align}
\vspace{-0.2cm}
\item  the periods around $\mA$ cycles on the boundary of the fundamental domain $D$ in figure \ref{fig:1} for any $y$ in the interior of $D$ are given in terms of Bernoulli numbers $B_r$ by, 
\bea
\label{1.per.g}
\oint _{\mA^K}  g^{I_1 \cdots I_r} {}_J (t,y) = (-2\pi i)^r { B_r \over r!} \, \delta ^{I_1 \cdots I_r K}_J
\eea
\end{enumerate}
Throughout, we will reserve the letter $t$ for integration variables, as we already did in (\ref{1.per.g}).

\vspace{\mylength}
%%%%%%%%%%%%%%%%%%%%%%%%%%%%%%%%%%%%%%%%%%%%%%%%%%%%%%%%%
%%%%%%%%%%%%%%%%%%%%%%%%%%%%%%%%%%%%%%%%%%%%%%%%%%%%%%%%%
\section{Lowest rank case}
\label{sec:3}
%%%%%%%%%%%%%%%%%%%%%%%%%%%%%%%%%%%%%%%%%%%%%%%%%%%%%%%%%
%%%%%%%%%%%%%%%%%%%%%%%%%%%%%%%%%%%%%%%%%%%%%%%%%%%%%%%%%
\vspace{\myotherlength}

In this section, we shall express the simplest non-trivial Enriquez kernel, namely  $g^I{}_J(x,y)$,  in terms of a convolution integral over $\mA$ homology cycles with the help of the prime form $E(x,y)$. 
One of the key results of this paper is the following theorem.

\sm

\noindent
{\bf Theorem 1.} 
\label{3.thm:1} 
\textit{The Enriquez kernel  $g^I{}_J(x,y)$ is given by,
\beq
\label{3.thm.1}
g^I{}_J(x,y)  =    \oint _{\mA^I} \om_J(t) \, \p_x \ln { E(x,y) \over E(x,t)}   -  \pi i \, \delta ^I_J \, \om_J(x)  
\eeq
with $x,y$ in the interior of $D$ and the cycle $\mA^I$ on the boundary of $D$ as in figure \ref{fig:1}.}

\sm

Before proving Theorem 1, we recall that the prime form $E(x,y)$ is defined by \cite{Fay:1973}, 
\bea
\label{prime}
E(x,y) = { \tet [\nu] ( \int^x_y\omega_I ) \over h_\nu(x) \, h_\nu (y) }
\eea
where the Riemann  function $\tet [\nu](\zeta_I)$ is evaluated on the Abel map $\zeta_I= \int^x_y\omega_I$ and $h_\nu$ is the holomorphic $(\half,0)$ form for odd spin structure $\nu$ defined up to a sign~by
\bea
h_\nu(x)^2 = \omega_K(x) \frac{\partial}{\partial \zeta_K } \tet[\nu](\zeta_I) \, \big|_{\zeta_I = 0}
\eea
 The prime form $E(x,y)$ is a multiple-valued holomorphic $(-\half, 0)$ form in $x, y \in \Sigma$ satisfying $E(y,x) = - E(x,y)$, has a simple zero at $x=y$ and is independent of $\nu$ thanks to the Riemann vanishing theorem for the $\tet$-function \cite{Fay:1973}.  The monodromies of $\p_x \ln E(x,y)$ around $\mA^L$ cycles vanish, while around $\mB_L$ cycles they are given by (see for example \cite{Fay:1973, DHoker:1988pdl}), 
\begin{align}
\label{sigmamonod}
\p_x \ln E(\mB_L \cdot x, y) & =   \p_x \ln  E(x,y)  - 2 \pi i  \om_L (x)
\no \\
\p_x \ln E(x, \mB_L \cdot y) & =   \p_x \ln  E(x,y)  + 2 \pi i  \om_L (x)
\end{align}
The derivative $\p_x$ is unchanged under the monodromy transformations assuming that the curves $\mA^L$ and $\mB_L$ are chosen to be continuously differentiable, as we shall do throughout.
Theorem 1 may equivalently be formulated directly in terms of the Riemann $\tet$-function for an arbitrary odd spin structure $\nu$ by using (\ref{prime}) to  replace $E(x,y)$ and $E(x,t)$ by $\tet[\nu]( \int^x_y\omega_I)$ and $\tet[\nu]( \int^x_t\omega_I )$, respectively.

\sm

\begin{figure}[htb]
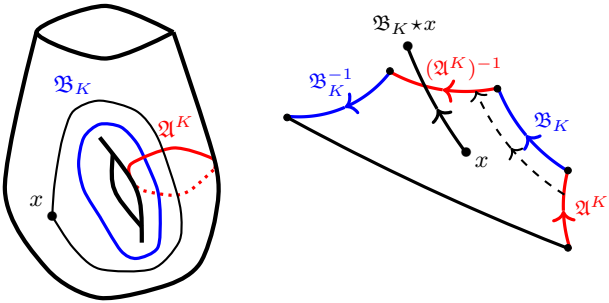

\begin{center}
\tikzpicture[scale=0.7]
\scope[xshift=1.5cm,yshift=0cm, scale=1, rotate=90]
% handles

% cycles
\draw[very thick, color=blue] plot [smooth] coordinates {(2,2) (2.5,2.3) (3, 2.4) (3.5, 2.3) (3.75, 2) (3.7,1.6) (3.4, 1.3)
(3, 1.1) (2.5, 0.9) (2, 0.85) (1.4, 0.9) (1.2, 1.3) (1.35, 1.6) (2,2)};
\draw[very thick, color=red] plot [smooth] coordinates { (2.7,1.4) ( 3, 1.4) (3.3, 0.6) (3.1, -0.14) (2.9, -0.18)};
\draw[very thick, color=red, dotted] plot [smooth] coordinates { (2.7,1.4) (2.5,1.2) (2.4,0.6) (2.5, 0.275) (2.9, -0.18)};
%\draw[very thick, color=red] plot [smooth] coordinates { (2.9,-0.18) (4,0.7) (5,2)};
% contour
\draw[ultra thick] plot [smooth] coordinates 
{(5.5,0.6) (3, -0.2) (2,-0.2) (1.1,0.1) (0.7,0.7) (0.45,2) (1,3) (2.5, 3.8)  (5.5,3.2)  };
\draw[ultra thick] plot [smooth] coordinates 
{(5.5,0.6) (5,1.9) (5.5,3.2) (6,1.9) (5.5,0.6)  };
% labels
\draw [color=red] (3.7,0.55) node{\small $\mA^K$};
\draw [color=blue] (4.5,2.5) node{\small $\mB_K$};
\draw[ultra thick] (1.5,1.2) .. controls (2.5, 1.2) .. (3.5,2);
\draw[ultra thick] (1.8,1.2) .. controls (2.5, 1.8) .. (3.2,1.75);
% base point 
%\draw (5,2) node{$\bullet$};
%\draw (5,2) [fill=black] circle(0.05cm) ;
\draw[thick, color=black] plot [smooth] coordinates {(2,2.9) (1.4, 2.5) (1,2) (1,1)  (1.5,0.5) (2, 0.45) (2.5,0.55) (3.5, 0.9) (4, 1.3)
(4.2, 2) (4, 2.5) (3.5, 2.7) (3, 2.8) (2.5, 2.87) (2,2.9)};
%\draw (5,2.38) node{ $\qq$};
\draw (2,2.9) node{ $\bullet$};
\draw (2.3,3.2) node{ $x$};
%\draw (7,2.95) node{ $\bullet$};
%\draw (7.4,2.97) node{ $y$};
\endscope
\scope[xshift=6cm,yshift=2.2cm, scale=4.2]
\draw  [very  thick, color=red, <-, domain=182:198] plot ({cosh(0.6)*cos(0)+sinh(0.6)*cos(\x)}, {cosh(0.6)*sin(0)+sinh(0.6)*sin(\x)});
\draw  [very  thick, color=red, domain=166:182] plot ({cosh(0.6)*cos(0)+sinh(0.6)*cos(\x)}, {cosh(0.6)*sin(0)+sinh(0.6)*sin(\x)});
\draw  [very  thick, color=red, <-, domain=261.5:282] plot ({cosh(0.65)*cos(85)+sinh(0.65)*cos(\x)}, {cosh(0.65)*sin(85)+sinh(0.65)*sin(\x)});
\draw  [very  thick, color=red,  domain=241:261.5] plot ({cosh(0.65)*cos(85)+sinh(0.65)*cos(\x)}, {cosh(0.65)*sin(85)+sinh(0.65)*sin(\x)});
\draw  [very  thick, color=blue, <-, domain=220:239] plot ({cosh(0.7)*cos(40)+sinh(0.7)*cos(\x)}, {cosh(0.7)*sin(40)+sinh(0.7)*sin(\x)});
\draw  [very  thick, color=blue, domain=201:220] plot ({cosh(0.7)*cos(40)+sinh(0.7)*cos(\x)}, {cosh(0.7)*sin(40)+sinh(0.7)*sin(\x)});
\draw  [very  thick, color=blue, <-, domain=293.5:323] plot ({cosh(0.5)*cos(125)+sinh(0.5)*cos(\x)}, {cosh(0.5)*sin(125)+sinh(0.5)*sin(\x)});
\draw  [very  thick, color=blue, domain=264:293.5] plot ({cosh(0.5)*cos(125)+sinh(0.5)*cos(\x)}, {cosh(0.5)*sin(125)+sinh(0.5)*sin(\x)});

\draw  [thick, color=black, dashed, ->, domain=238:220] plot ({cosh(0.9)*cos(40)+sinh(0.9)*cos(\x)}, {cosh(0.9)*sin(40)+sinh(0.9)*sin(\x)});
\draw  [thick, color=black, dashed, ->, domain=220:203] plot ({cosh(0.9)*cos(40)+sinh(0.9)*cos(\x)}, {cosh(0.9)*sin(40)+sinh(0.9)*sin(\x)});

\draw  [very thick, color=black, ->, domain=217:210] plot ({cosh(1.45)*cos(40)+sinh(1.45)*cos(\x)}, {cosh(1.45)*sin(40)+sinh(1.45)*sin(\x)});
\draw  [very thick, color=black,  domain=210:201] plot ({cosh(1.45)*cos(40)+sinh(1.45)*cos(\x)}, {cosh(1.45)*sin(40)+sinh(1.45)*sin(\x)});

\draw  [very thick, color=black,  domain=241.3:248.6] 
plot ({cosh(3.1)*cos(65.5)+sinh(3.1)*cos(\x)}, {cosh(3.1)*sin(65.5)+sinh(3.1)*sin(\x)});
\draw (0.57,0.16) [fill=black] circle(0.015cm) ;
\draw (0.251,0.53) [fill=black] circle(0.015cm) ;
\draw (-0.7,0.4) [fill=black] circle(0.015cm) ;
\draw (-0.235,0.61) [fill=black] circle(0.015cm) ;
\draw (0.57,-0.19) [fill=black] circle(0.015cm) ;
\draw [color=red] (0.68,0) node{\footnotesize $\mA^K$};
\draw [color=blue] (0.5,0.38) node{\footnotesize $\mB_K$};
\draw [color=red] (0.1,0.62) node{\footnotesize $(\mA^K)^{-1}$};
\draw [color=blue] (-0.5,0.56) node{\footnotesize $\mB_K^{-1}$};
%\draw (0,0) node{{\Large $D$}};
\draw (0.11,0.24) node{{$\bullet$}};
\draw (0.18,0.2) node{{$x$}};
\draw (-0.154,0.72) node{{$\bullet$}};
\draw (-0.17,0.82) node{{\footnotesize $\mB_K \! \cdot \! x$}};
\endscope
%%%
%%%

\endtikzpicture
\caption{In the left panel, the point $x$ in one handle part of $\Sigma$ is moved along the cycle $\mB_K$, intersecting  the cycle $\mA^K$ once, and returning to $x$. In the right panel, the same move is shown in one handle part of the fundamental domain $D$.
 \label{fig:2}}
\end{center}
\end{figure}

\begin{proof}
To prove Theorem 1, we consider the combination,
\bea
\label{3.gE}
  g^I{}_J(x,y)  -  \oint _{\mA^I} \om_J(t) \, \p_x \ln { E(x,y) \over E(x,t)}
\eea
The pole of $g^I{}_J(x,y)$ in $x$ at $y$ with residue $\delta^I_J$  is cancelled by the pole of $\delta ^I_J \, \p_x \ln E(x,y)$ so that the combination in (\ref{3.gE}) is holomorphic in $x$ and $y$. Its monodromies in $x,y$ around $\mA$ cycles vanish, while its monodromies in $y$ around $\mB$ cycles vanish by combining (\ref{2.mongy}) with the second line in (\ref{sigmamonod}).  To evaluate the monodromy in $x$ around $\mB$ cycles of (\ref{3.gE}), we need to take into account two effects, namely,
\begin{itemize}
\itemsep 0in
\item[(i)] the monodromy of the integrand as $x \to \mB_L \cdot x$;
\item[(ii)] the contribution from the pole in the variable $t$ at $x$ as $x \to \mB_L \cdot x$ crosses the $\mA^I$ cycle in the formulas of (\ref{3.thm.1}) and (\ref{3.gE}), as illustrated in figure \ref{fig:2}.
\end{itemize}
The contribution from (i) vanishes because the $\mB$ monodromies (\ref{sigmamonod}) of the prime forms cancel from the ratio in (\ref{3.gE}).   The contribution of (ii) is given by the residue of the integrand in $t$ at $x$ which by $\mJ(\mA^I, \mB_L) = \delta ^I_L$  evaluates to $ - 2 \pi i \delta ^I_L \om _J(x)$ and cancels the monodromy of $g^I{}_J(x,y)$ in $x$ given in (\ref{2.mongx}).  Accordingly, the $\mA^I$ integral over $ \p_x \ln   E(x,t)$ in (\ref{3.thm.1}) creates a branch cut in the variable $x\in \Sigma$ as expected for $g^{I}{}_J(x,y)$.

\sm

Putting all together we conclude that (\ref{3.gE}) is independent of $y$ since it is a single-valued holomorphic scalar in $y \in \Sigma$ and that it must be a  linear combination of the $\om_K(x)$ forms since it is a single-valued holomorphic $(1,0)$ form in $x \in \Sigma$. The corresponding coefficients are fixed by imposing the $\mA$ periods of (\ref{1.per.g}) at $r=1$, which gives the last term of (\ref{3.thm.1}) and completes the proof of Theorem~1. Note that the $\mA^K$ period of $\oint_{\mA^I} \omega_J(t) \p_x \ln \frac{E(x,y)}{E(x,t)}$ with respect to $x$ evaluates to $2\pi i \delta^{IK}_J$ since the point $x$ in the integrand of (\ref{3.thm.1}) is in the interior of $D$ in figure \ref{fig:1} and must be moved across the contour for $t$ to make the expression (\ref{1.per.g}) for the $\mA$-period applicable.
\end{proof}

By decomposing $g^I{}_J(x,y) = \varpi ^I{}_J(x) - \delta ^I_J \chi(x,y)$ into its trace part $\chi(x,y)$ and its $y$-independent traceless part $\varpi ^I{}_J(x)$ subject to $\varpi ^K{}_K(x)=0$,  we readily obtain an expressions for the trace,
\bea
\label{3.cor.1}
\chi(x,y) = -{ 1 \over h} \p_x \ln \Big ( E(x,y)^h \sigma(x) \Big )  + {\pi i \over h} \sum _{K=1} ^h \om_K(x)
\eea
Here, the Fay form $\sigma (x)$ is a multiple-valued, holomorphic, nowhere vanishing $( {h \over 2}, 0)$ form defined in  \cite{Fay:1973} by,
\bea
\label{Fay}
\ln \sigma (x) =  - \sum _{K=1}^h \oint _{\mA^K} \om_K(t) \ln E(x,t) 
\eea

\vspace{\mylength}

%%%%%%%%%%%%%%%%%%%%%%%%%%%%%%%%%%%%%%%%%%%%%%%%%%%%%%%%%
%%%%%%%%%%%%%%%%%%%%%%%%%%%%%%%%%%%%%%%%%%%%%%%%%%%%%%%%%
\section{Enriquez kernels via convolution over $\mA$-cycles}
\label{sec:5}
%%%%%%%%%%%%%%%%%%%%%%%%%%%%%%%%%%%%%%%%%%%%%%%%%%%%%%%%%
%%%%%%%%%%%%%%%%%%%%%%%%%%%%%%%%%%%%%%%%%%%%%%%%%%%%%%%%%
\vspace{\myotherlength}

In this section, we derive the advertised new representations of Enriquez kernels as $\mA$ convolutions over Enriquez kernels of lower rank from the Fay identity \cite{DHoker:2024ozn}, 
\begin{align}
0& = 
\Big (  g^{ J}{}_K  (y,t)  - g^{J}{}_K(y,z)   \Big ) g^{I_1 \cdots I_r } {}_J  (t,z)  
\label{13.a} \\ &\quad
+ \sum_{k = 0}^r g^{  I_1 \cdots I_k } {}_J(t,y) \, g^{J I_{k + 1} \cdots I_r}{}_K(y,z)
\no \\ &\quad
+ \om_J(y) \Big[  g^{ I_1 \cdots I_r J} {}_K(t,y) +  g^{(J \shuffle I_1 \cdots I_{r-1}) I_r} {}_K(t,z) \Big] \notag
\end{align}
proven in appendix A (see for instance \cite{DHoker:2024ozn} for the shuffle product $\shuffle$, and  \cite{Baune:2024ber} for a more formal proof of (\ref{13.a}).
%; we use Einstein summation conventions
%for repeated indices such as $J$ in (\ref{13.a}) and (\ref{4.thm.2}) below). 
The resulting recursion relations of Enriquez kernels are the content of the following theorem.

\sm

\begin{figure}[htb]
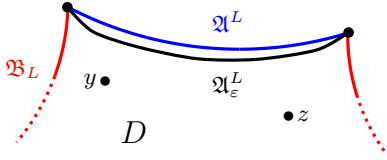

\begin{center}
\tikzpicture[scale=1]
\scope[xshift=6cm,yshift=2.2cm, scale=4.2, rotate=0]
\draw  [very  thick, color=red,  domain=179:198] plot ({cosh(0.6)*cos(20)+sinh(0.6)*cos(\x)}, {cosh(0.6)*sin(20)+sinh(0.6)*sin(\x)});
\draw  [very  thick, color=red,  dotted, domain=215:198] plot ({cosh(0.6)*cos(20)+sinh(0.6)*cos(\x)}, {cosh(0.6)*sin(20)+sinh(0.6)*sin(\x)});
\draw  [very  thick, color=red,  domain=359:340] plot ({cosh(0.65)*cos(155)+sinh(0.65)*cos(\x)}, {cosh(0.65)*sin(155)+sinh(0.65)*sin(\x)});
\draw  [very  thick, color=red,  dotted, domain=340:320] plot ({cosh(0.65)*cos(155)+sinh(0.65)*cos(\x)}, {cosh(0.65)*sin(155)+sinh(0.65)*sin(\x)});
\draw  [very  thick, color=blue,  domain=287:270] plot ({cosh(1)*cos(85)+sinh(1)*cos(\x)}, {cosh(1)*sin(85)+sinh(1)*sin(\x)});
\draw  [very  thick, color=blue, domain=270:242] plot ({cosh(1)*cos(85)+sinh(1)*cos(\x)}, {cosh(1)*sin(85)+sinh(1)*sin(\x)});
\draw[very thick] plot [smooth] coordinates  {(0.48,0.415) (0.38, 0.36) (0.19, 0.33) (0, 0.332) (-0.19, 0.37) (-0.33, 0.418) (-0.41,0.495) };

\draw (0.48,0.415) [fill=black] circle(0.015cm) ;
\draw (-0.41,0.495) [fill=black] circle(0.015cm) ;
%\draw (-0.562,0.08) [fill=black] circle(0.015cm) ;
%\draw (0.585,0.05) [fill=black] circle(0.015cm) ;

\draw [color=red] (-0.55,0.3) node{\small  $\mB_L$};
\draw [color=blue] (0.1,0.45) node{\small $\mA^L$};
%\draw [color=red] (0.75,0.3) node{\footnotesize $(\mB_{L-1})^{-1}$};
\draw  (0.1,0.25) node{\small $\mA^L_\ep$};
\draw  (-0.2,0.1) node{ \large $D$};
\draw (-0.29,0.26)node{$\bullet$}node[left]{$y$};
\draw (0.29,0.15)node{$\bullet$}node[right]{$z$};
\endscope
%%%
%%%

\endtikzpicture
\caption{For $\ep >0$, the cycle $\mA^L_\ep$ is a small deformation of $\mA^L$ that is homotopic to $\mA^L$ and contained in the interior of $D$. 
 \label{fig:3}}
\end{center}
\end{figure}

{\bf Theorem 2.}  
\textit{The Enriquez kernels satisfy the following recursion relations in $r\geq 1$ for $y,z$ in the interior of~$D$ 
and cycles $\mA^L$ on the boundary of $D$ in figure \ref{fig:1},}
\begin{align}
\label{4.thm.2}
&g^{L I_1 \cdots I_r}{}_K(y,z)  =  
- \oint _{\mA^L}  g^{ J}{}_K  (y,t)  g^{I_1 \cdots I_r } {}_J  (t,z)  
\no  \\ &\quad\quad
- \sum_{k = 1}^{r-1} (-2 \pi i)^k { B_k \over k!} \, \delta ^{I_1 \cdots I_k }_L \, g^{L I_{k + 1} \cdots I_r}{}_K(y,z)
\no \\ &\quad\quad
-   \om_K(y)   (-2 \pi i)^{r+1} { B_{r+1} \over r!} \, \delta ^{I_1 \cdots I_r  L}_K
\end{align}
\textit{Here, and throughout the remainder of this letter,
 the convolution integral over $\mA^L$ in the first line of (\ref{4.thm.2}) is to be defined as a limit,}
\bea
\label{limit}
 \lim _{\ep \to 0}  \oint _{\mA^L_\ep }  g^{ J}{}_K  (y,t)  g^{I_1 \cdots I_r } {}_J  (t,z)   
 \label{eplim}
\eea
\textit{where $\mA^{L} _\ep$ is a cycle that is homotopic to $\mA^{L}$ but displaced by $\ep > 0$  from the boundary of $D$ into the interior of $D$ without crossing $y,z \in D$,  as shown in figure \ref{fig:3}. In this way, the integration of $t$ over $\mA^{L}_\ep $ in (\ref{limit}) is contained in the interior of $D$ where the integrand is defined. }

\sm

\begin{proof}
The proof of Theorem 2 proceeds by solving (\ref{13.a}) for the $k=0$ term of the sum in the second line, integrating $t$ over the cycle $\mA^L$ and using the $\mA$ periods (\ref{1.per.g}) of the Enriquez kernels. 
\end{proof}

\sm

In appendix B, the monodromies of (\ref{4.thm.2}) as $y$ is taken around a $\mB$ cycle are shown to reproduce the defining property (\ref{2.mongx}) of Enriquez kernels. This matching involves subtle contributions arising from the pole in $t$ at $y$ as $t$ crosses the branch cut furnished by the $\mA^L$ cycle of integration.
Appendix B also illustrates why the poles $\partial_x \ln E(x,y) = (x{-}y)^{-1}+ {\rm reg}$ of the prime forms entering the integral representation (\ref{3.thm.1}) of $g^J{}_K(y,t)$ necessitate the $\ep$ displacement of the integration contour and the limit $\ep \rightarrow 0$ as specified in Theorem~2.
In a nutshell, deforming the contour $\mA^L_\ep$ of figure \ref{fig:3} to be displaced outside instead of inside $D$ would cross the integration contour in (\ref{3.thm.1}) on $\partial D$ and would lead to non-trivial residue contributions from the prime forms as exemplified in (\ref{B.9}).

% which also illustrate the necessity of the $\ep$ displacement and the limit $\ep \rightarrow 0$ as specified in Theorem~2.

\vspace{\mylength}

\subsection{Closure under convolution over $\mA$ cycles}

\vspace{\myotherlength}

In addition to the statement of Theorem 2, one may reformulate (\ref{4.thm.2}) as showing that the products $g^{I_1 \cdots I_r}{}_J(t,z)$ with $g^J{}_K(y,t)$ close under convolution over an arbitrary $\mA$ cycle. Generalizing this result to the convolution of two arbitrary Enriquez kernels gives the following theorem. 

\sm

{\bf Theorem 3.} 
\label{4.thm:3}
\textit{The convolution over an arbitrary cycle $\mA^L$  of the product of two arbitrary Enriquez kernels 
with $r,s\geq 0$ closes in the sense that it produces a linear combination of Enriquez kernels, 
\begin{align}
& \oint_{\mA^L} \,  g^{P_1 \cdots P_s A}{}_B  (y,t) \, g^{I_1 \cdots I_r} {}_K  (t,z) 
\label{closeconv}  \\ &
\ = 
(-2 \pi i )^r {B_r \over r!}  \, \delta ^{I_1 \cdots I_r   L}_K\, g ^{P_1 \cdots P_s A}{}_B(y,z) 
\no \\ &
\quad +  \delta^A_B\sum_{\ell=0}^s  \bigg\{
R_{r,s+1-\ell}   \delta ^{P_s \cdots P_{\ell+1}  I_1 \cdots I_rL}_K  g^{P_1 \cdots P_\ell}{}_L(y,z) 
 \notag \\
 &\quad \quad  - \sum_{k = 0}^r 
R_{k, s-\ell} \delta ^{P_s \cdots P_{\ell+1}  I_1 \cdots I_k }_L
g^{P_1\cdots  P_\ell L I_{k + 1} \cdots I_r}{}_{ \! K}(y,z) 
\no\bigg\}
\end{align}
where we have set $R_{m,n} = (2 \pi i)^{m+n} (-)^m B_{m+n}/(m! \, n!)$.}

\sm

\begin{proof}
The proof of all cases with $r\geq 1$ proceeds from the general contracted Fay identity for arbitrary Enriquez kernels of Conjecture 9.7 in \cite{DHoker:2024ozn}, which was proven in \cite{Baune:2024ber}, by first obtaining the uncontracted Fay identity in section 9.3.2 of \cite{DHoker:2024ozn} and then integrating this identity over the $\mA^L$ cycle. The case $r=0$ of (\ref{closeconv}) follows from integrating Corollary 9.3 of \cite{DHoker:2024ozn} over $x$.
\end{proof}

\sm

We note that the statement (\ref{4.thm.2}) of Theorem 2 follows from
(\ref{closeconv}) of Theorem 3 upon specializing $s=0$ and contracting with $\delta^K_A$. Still, iterative
use of (\ref{4.thm.2}) suffices to express arbitrary Enriquez kernels as convolutions of
the Abelian differentials and prime forms in (\ref{3.thm.1}), a property that is most clearly exhibited in Theorem 2.

\vspace{\mylength}

\subsection{Genus one case}

\vspace{\myotherlength}

For genus one, the Enriquez kernels reproduce the meromorphic Kronecker-Eisenstein coefficients $g^{(r)}$ \cite{Enriquez:2011},
\beq
g^{I_1 \cdots I_r}{}_J(x,y) \, \big|_{h=1}
= dx \, g^{(r)}(x{-}y) \, , \ \ \ \ \ \ r\geq 1
\eeq
The multi-valued functions $g^{(r)}$ on the torus have found widespread use as integration kernels for elliptic polylogarithms \cite{Broedel:2017kkb, Bourjaily:2022bwx} and can be expressed in terms of the  odd Jacobi theta function $\vartheta_1$ by Laurent expansion in $\eta$,
\beq
\frac{ \vartheta'_1(0) \vartheta_1(z{+}\eta) }{\vartheta_1(z)  \vartheta_1(\eta) }
= \frac{1}{\eta}+ \sum_{r=1}^{\infty}\eta^{r-1 }g^{(r)}(z)
\label{KEser}
\eeq
Hence, the $\mA$ integrals over prime forms in the expressions (\ref{3.thm.1}), (\ref{4.thm.2}) for Enriquez kernels at genus $h$ can be explicitly performed at genus $h=1$ where $\omega_I(x) |_{h=1}= dx$ and,
\beq
\partial_x \ln E(x,y) \, \big|_{h=1} =
g^{(1)}(x{-}y) = \partial_x \ln \vartheta_1(x{-}y)
\eeq
and lead to the combinations of  $\vartheta_1$ generated by (\ref{KEser}). Conversely, our construction of Enriquez kernels provides recursive integral representations at $r\geq 1$,
\begin{align}
&g^{(r+1)}(y{-}z)  =  
- \int^1_0 dt \, g^{(1)} (y{-}t)  g^{(r)} (t{-}z)  
\label{h1rec} \\ 
 &\quad
  -     (-2 \pi i)^{r+1} { B_{r+1} \over r!}  
- \sum_{k = 1}^{r-1} (-2 \pi i)^k { B_k \over k!} \, g^{(r+1-k)}(y{-}z)
\notag
\end{align}
which follow from (\ref{4.thm.2}) and are valid for $y,z$ in the interior of the fundamental domain for the torus, namely for $0< \Im y, \, \Im z  <  \Im \Omega_{11}$. The recursion relation  (\ref{h1rec}) for meromorphic genus-one kernels $g^{(r)}$ can be directly proven based on the Fourier decomposition of their single-valued counterparts $f^{(r)}$ \cite{BrownLevin, Broedel:2014vla} and, as far as we know, has not earlier appeared in print.

\vspace{\mylength}
%%%%%%%%%%%%%%%%%%%%%%%%%%%%%%%%%%%%%%%%%%%%%%%%%%%%%%%%%
%%%%%%%%%%%%%%%%%%%%%%%%%%%%%%%%%%%%%%%%%%%%%%%%%%%%%%%%%
\section{Varying moduli}
\label{sec:6}
%%%%%%%%%%%%%%%%%%%%%%%%%%%%%%%%%%%%%%%%%%%%%%%%%%%%%%%%%
%%%%%%%%%%%%%%%%%%%%%%%%%%%%%%%%%%%%%%%%%%%%%%%%%%%%%%%%%
\vspace{\myotherlength}

In this section, we evaluate the variational derivatives $\delta_{ww}$ and $\delta_{\bar w \bar w}$ of the Enriquez kernels at an arbitrary point $w \in \Sigma$. (The notation is standard and not to be confused with the one for the Kronecker $\delta$-symbol.) In conformal field theory, $\delta_{ww}$ and $\delta_{\bar w \bar w}$ correspond to the holomorphic and anti-holomorphic parts of the traceless stress tensor,  $T_{ww}$  and $T_{\bar w \bar w}$, respectively. The variational derivatives  provide an efficient tool to evaluate the dependence on the complex structure moduli of the Riemann surface $\Sigma$. The derivative with respect to a holomorphic modulus $m$ of an arbitrary  function $X$ is given by
(see for example \cite{Verlinde:1986kw,DHoker:1988pdl}), 
\bea
\label{pX}
{ \p X \over \p m} = { 1 \over 2 \pi} \int_\Sigma \mu_{\bar w} {}^w \big ( \delta_{ww} X \big ) 
\eea
where $\mu$ is the Beltrami differential associated with the modulus $m$. For example,  the local  variational derivative of the period matrix is given by,
\bea
\delta _{ww} \Omega _{IJ} = 2 \pi i  \, \om_I(w) \om_J(w) 
\label{donom}
\eea
Among  the $\half h (h+1)$ holomorphic quadratic differentials $\om_I(w) \om_J(w)$, only $3h-3$ are linearly independent for $h \geq 2$ (only 1 for $h=1$) and generate independent  variations of the moduli of $\Sigma$.   Since we have $\delta _{\bar w \bar w} \Omega_{IJ}=0$ and the kernels  $g^{I_1 \cdots I_r}{}_J(x,y) $ are locally holomorphic in moduli by construction, their $\delta_{\bar w\bar w}$ variation vanishes.

\sm

The $\delta _{ww}$ variations of Enriquez kernels, to be considered  in the sequel, will be crucial to obtain differential equations in moduli for modular graph tensors \cite{Kawazumi:lecture, DHoker:2020uid} and for higher-genus analogues \cite{Baune:2024} of elliptic multiple zeta values \cite{Enriquez:Emzv} from their integral representations (see for instance \cite{DHoker:2014oxd} for the derivation of a Laplace equation for the genus two Kawazumi-Zhang invariant via (\ref{pX}) that has important applications in string amplitudes). 

\sm

The basic variational formulas for functions that are needed here are as follows (see for example \cite{Verlinde:1986kw,DHoker:1988pdl}), 
\bea
\label{14.a}
\delta _{ww} \, \p_x & = & \p_x \delta_{ww} 
\\
\delta _{ww} \, \om_I(x) & = & \om_I(w) \p_w \p_x \ln E(w,x)
\no \\
\delta _{ww} \ln E(x,y) & = & - \thalf \big ( \p_w \ln E(w,x) - \p_w \ln E(w,y) \big )^2
\no 
\eea
Equations (\ref{donom}) and (\ref{14.a}) together with Leibniz's rule for $\delta _{ww}$ provide sufficient information to evaluate the variational derivatives of the Enriquez kernels.

\sm

The variational derivative for rank $r=1$ is obtained using the formula 
for $g^I{}_J(x,y)$ given in (\ref{3.thm.1}) and (\ref{14.a}), 
\begin{align}
\label{14.c}
&\delta _{ww} \, g^I{}_J(x,y)  =   \p_w \p_x \ln E(w,x)  \, g^I{}_J(w,y)
 \\ &\quad \quad
-   \om_J(w) \, \oint _{\mA^I}   \p_x \ln  E(x,t)    \,  \p_w \p_t \ln E(w,t) 
\no
\end{align}
The second line can be rewritten as
\bea
\label{13.h}
  \oint  _{\mA^I} \p_x \ln E(x,t) \,  \p_w \p_t \ln E(w,t)  = \p_w \chi^I (x,w) 
\eea
using the convolution representation (\ref{4.thm.2}) of $\chi^I (x,w)  = - \frac{1}{h} \delta^L_K g^{IK}{}_L(x,w) $ and noting the additional $w$-derivative in (\ref{14.c}). Substituting this result into (\ref{14.c}), and using $\p_w \chi^I(x,w) \delta ^M_J = - \p_w g^{IM}{}_J(x,w)$ we obtain, 
\begin{align}
\label{14.fin.2}
\delta _{ww} \, g^I{}_J(x,y) & =   \p_w \p_x \ln E(w,x)  \, g^I{}_J(w,y)  
\notag \\
&\quad +  \om_M(w) \, \p_w g^{IM}{}_J (x,w)  
\end{align}
The first term on the right side is mandated by the fact that $g^I{}_J(x,y)$ is a $(1,0)$ form in $x$, as is familiar from the diffeomorphism Ward identities in conformal field theory (see for example \cite{Verlinde:1986kw,DHoker:1988pdl}).

\sm

To obtain the variational derivative for arbitrary rank $r \geq 1$, we start with the recursion relations (\ref{4.thm.2}) and obtain the result stated in the following theorem.

\sm

{\bf Theorem 4.}
\textit{The variational derivative of the Enriquez kernel $g^{I_1 \cdots I_r}{}_J(x,y)$ is given by,}
\begin{align}
\delta_{ww}  \, &g^{I_1 \cdots I_r}{}_J(x,y) = \p_w \p_x \ln E(w,x) g^{I_1 \cdots I_r}{}_J(w,y) \notag \\
&+ \sum_{k=1}^r \p_w g^{I_1 \cdots I_k M}{}_J(x,w) \, g^{I_{k+1} \cdots I_r}{}_M(w,y)
\label{5.thm.1}
\end{align}
The proof of this theorem is presented in appendix C.

\vspace{\mylength}
%%%%%%%%%%%%%%%%%%%%%%%%%%%%%%%%%%%%%%%%%%%%%%%%%%%%%%%%%
%%%%%%%%%%%%%%%%%%%%%%%%%%%%%%%%%%%%%%%%%%%%%%%%%%%%%%%%%
\section{Parallels  with single-valued kernels}
%%%%%%%%%%%%%%%%%%%%%%%%%%%%%%%%%%%%%%%%%%%%%%%%%%%%%%%%%
%%%%%%%%%%%%%%%%%%%%%%%%%%%%%%%%%%%%%%%%%%%%%%%%%%%%%%%%%
\vspace{\myotherlength}

A striking correspondence has emerged between the relations among meromorphic multiple-valued Enriquez kernels $g^{I_1 \cdots I_r}{}_J(x,y)$ and those among non-meromorphic single-valued modular tensors $f^{I_1 \cdots I_r}{}_J (x,y)$ of \cite{DHoker:2023vax,DHoker:2023khh}. This correspondence was already brought to light in \cite{DHoker:2024ozn} where the Fay identities in both cases were found to be related by a simple substitution $f^{I_1 \cdots I_r}{}_J(x,y) \leftrightarrow g^{I_1 \cdots I_r}{}_J(x,y)$ and in \cite{DHoker:2025szl} where the associated flat connections were related through a gauge transformation together with a Lie algebra automorphism. 

\sm

In the present paper we find additional elements of the correspondence. A first is between the recursion relations of Theorem 2 and those of the $f$ kernels for $r \geq 2$ \cite{DHoker:2023vax}, 
\begin{align}
f^{L I_1 \cdots I_r}{}_{\! K}(x,y)  =   \tfrac{i}{2} \! \int _\Sigma \! f^J {}_{\! K} (x,t)  \bar \om^{L} (t)  f^{I_1 \cdots I_r}{}_{\! J}(t,y)
\label{recfs}
\end{align}
Apart from the correction terms on the second and third lines of (\ref{4.thm.2}) involving lower-rank Enriquez kernels, the correspondence consists of replacing the integral over $\mA^L$ by an integral of $\bar \om^L$ on $\Sigma$. Moreover, the initial condition for the recursion (\ref{recfs}),
\beq
f^I{}_J(x,y) =   \tfrac{i}{2} \! \int _\Sigma \!  \bar \om^{I} (t) \om_{J}(t) \, \p_x \big[ \cG(x,y) {-} \cG(x,t) \big]
\label{frk2}
\eeq
matches the representation (\ref{3.thm.1})  of $g^I{}_J(x,y)$ under this dictionary between integrals $\oint_{\mA^L}$ and $\int_\Sigma \bar \omega^L$, up to the holomorphic term $\sim \omega_J(x)$, provided we trade ${-}\ln E(x,z)$ for the Arakelov Green function $\cG(x,z)$  (see e.g.\ \cite{Falt, DHoker:2017pvk}).

\sm

A further element of the correspondence is observed between the moduli variation relations of Theorem 4 and those of the $f$ kernels, 
\begin{align}
\delta_{ww} \, &f_{I_1 \cdots I_r}{}^J(x,y)  =  - \p_w \p_x  {\cal G}(w,x) f_{I_1 \cdots I_r}{}^J(w,y)
\notag \\
&+ \sum_{k=1}^r \p_w f_{I_1 \cdots I_k M}{}^J(x,w) 
f_{I_{k+1} \cdots I_r}{}^M(w,y)
\label{dwwfs}
\end{align}
where the indices have been lowered and raised with the help of the tensor $Y_{IJ}= (\Im \Omega)_{IJ}$ and its inverse, respectively. The proofs of (\ref{5.thm.1}) and (\ref{dwwfs}) proceed in close parallel to one another using the variational formulas of (\ref{14.a}) and suitable integrations by part in  the variable $t$.

\vspace{\mylength}
%%%%%%%%%%%%%%%%%%%%%%%%%%%%%%%%%%%%%%%%%%%%%%%%%%%%%%%%%
%%%%%%%%%%%%%%%%%%%%%%%%%%%%%%%%%%%%%%%%%%%%%%%%%%%%%%%%%
\section{Conclusion and outlook}
%%%%%%%%%%%%%%%%%%%%%%%%%%%%%%%%%%%%%%%%%%%%%%%%%%%%%%%%%
%%%%%%%%%%%%%%%%%%%%%%%%%%%%%%%%%%%%%%%%%%%%%%%%%%%%%%%%%
\vspace{\myotherlength}

In this paper, we have exhibited simple explicit recursion relations for the Enriquez kernels via which they may be expressed as multiple convolution integrals over homology cycles on $\Sigma$ with integrands composed of only the prime form and holomorphic Abelian differentials.
Our results provide both conceptual and computational links between different formulations of higher-genus polylogarithms with either meromorphic or modular covariant integration kernels and pave the way for combining the strengths of both approaches. We expect the new representations of Enriquez kernels to find a wide range of applications beyond the moduli variations, including,
\begin{itemize}
\item Studying the behavior of Enriquez kernels under degenerations of the Riemann surface $\Sigma$ using the results of \cite{Fay:1973} and adapting those of~\cite{DHoker:2017pvk, DHoker:2018mys}.
\item Assuming that good numerical control over the holomorphic Abelian differentials $\omega_I$ and the associated Abel map can be achieved at higher genus, we expect that our construction may offer a new and profitable line of attack for the numerical evaluation of Enriquez kernels: the only other ingredient in the integrand is the Riemann $\tet$ function, which is under excellent numerical control already.
\end{itemize}
As a first string-theory application, we shall show in a companion paper \cite{inprogress:2025} that cyclic products of Szeg\"o kernels admit systematic descent equations to a linear combination of tensor-like objects that are independent of the locations of the Szeg\"o kernels with coefficients given by Enriquez kernels. Moreover, since it is the meromorphic formulation of elliptic polylogarithms \cite{Levin:1997, Levin:2007, CEE, Broedel:2017kkb} that found most prominent appearance in Feynman-integral evaluations \cite{Bourjaily:2022bwx}, we expect our results to impact precision computations in particle physics and gravity, for instance by unravelling the role of Siegel modular forms for Feynman integrals as pioneered in \cite{Duhr:2024uid} or by offering a prototype for the construction of integration kernels~on higher-dimensional varieties from similar convolution integrals.

\sm

{\it Acknowledgements}: We are grateful to Benjamin Enriquez and Federico Zerbini for collaboration on related topics. Two anonymous referees are thanked for their valuable comments on an earlier version of this work. OS cordially thanks Federico Zerbini for valuable discussions. The research of ED is supported in part by NSF grant PHY-22-09700.  The research of OS is supported by the strength area ``Universe and mathematical physics'' which is funded by the Faculty of Science and Technology at Uppsala University.

\vspace{\mylength}
%%%%%%%%%%%%%%%%%%%%%%%%%%%%%%%%%%%%%%%%%%%
%%%%%%%%%%%%%%%%%%%%%%%%%%%%%%%%%%%%%%%%%%%
\section{Appendix A: Proof of the Fay identity}
\label{sec:A}
%%%%%%%%%%%%%%%%%%%%%%%%%%%%%%%%%%%%%%%%%%%
%%%%%%%%%%%%%%%%%%%%%%%%%%%%%%%%%%%%%%%%%%%
\vspace{\myotherlength}

In this appendix, we shall provide a proof of the Fay identity (\ref{13.a}) that is more direct than the proof given in~\cite{Baune:2024ber}. We begin by rephrasing (\ref{13.a}) as the vanishing of, 
\begin{align}
\label{13.alt}
&S^{I_1 \cdots I_r}{}_{\! K}(x,y,z)  = 
\Big (  g^{ J}{}_{\! K}  (y,x) \!  - \! g^{J}{}_{\! K}(y,z)   \Big ) g^{I_1 \cdots I_r } {}_J  (x,z)  
\no \\ &\quad
+ \sum_{k = 0}^r g^{  I_1 \cdots I_k } {}_J(x,y) \, g^{J I_{k + 1} \cdots I_r}{}_{\! K}(y,z)
 \\ &\quad
+ \om_J(y) \Big[  g^{ I_1 \cdots I_r J} {}_{\! K}(x,y) +  g^{(J \shuffle I_1 \cdots I_{r-1}) I_r} {}_{\! K}(x,z) \Big] \notag
\end{align}
For $r=0$, the identity $S^\emptyset {}_K(x,y,z)=0$ holds in view of the meromorphic interchange lemma of \cite{DHoker:2024ozn} applied to,
\bea
S^\emptyset {}_K(x,y,z) = \om_J  (x)  g^J{}_K  (y,x) + \om_J(y) g^J{}_K(x,y)
\eea
For $r \geq 1$,  it will be convenient to use the notation,
\bea
\Delta _L ^{(x)} f(x) = f(\mB_L \cdot x) - f(x)
\label{deltnot}
\eea
for an arbitrary form $f$ to prove the following lemma.

\sm

\noindent
{\bf Lemma 1.}
\textit{The forms $S^{I_1 \cdots I_r}{}_K(x,y,z) $ defined in (\ref{13.alt})
\begin{itemize}
\itemsep=-0.02in
\item[(a)] are holomorphic in $x,y,z$;
\item[(b)] have vanishing $\mA$ monodromies in $x,y,z$; 
\item[(c)] have the following $\mB$ monodromies in $x$ and $z$,
\begin{align}
\label{B.lem.1}
\Delta _L ^{(x)} S^{I_1 \cdots I_r}{}_K & (x, y,z) 
 =  \sum_{\ell=1}^r { (-2 \pi i)^\ell \over \ell!} \, \delta ^{I_1 \cdots I_\ell} _L
 \notag \\
 &\hspace{1.8cm} \times   \, S^{I_{\ell+1} \cdots I_r} {}_K(x,y,z)
\no \\
\Delta _L ^{(z)} S^{I_1 \cdots I_r}{}_K& (x, y,z)  = 
\sum_{\ell=1}^r {(2 \pi i)^\ell \over \ell!} \, S^{I_1 \cdots I_{r-\ell}} {}_L(x,y,z)
\notag \\
&\hspace{1.8cm}\times   \,  \delta ^{I_{r-\ell+1} \cdots I_{r-1} }_L \delta ^{I_r}_K
\end{align}
as well as vanishing $\mB$ monodromies in $y$. 
\end{itemize}}
%}

\noindent
\begin{proof} 
We shall give separate proofs for  parts \textit{(a), (b)} and~\textit{(c)}.

\sm

\textit{(a)} We consider $S^{I_1 \cdots I_r}{}_K(x,y,z) $ defined in (\ref{13.alt}) for $r \geq 1$ and $x,y,z$ in the  fundamental domain $D$ of figure~\ref{fig:1}, where  $g^{I_1 \cdots I_r}{}_J(x,y)$ is holomorphic for $r \geq 2$ and has a unique simple pole in $x$ at $y$ with residue $\delta^{I_1}_J$ for ${r=1}$. In the variable $x$ for $S^{I_1 \cdots I_r}{}_K(x,y,z) $ at $r \geq 2$, the residues of the simple poles at $y$ in the first term inside the parentheses on the first line, and  from the $k=1$ contribution on the second line cancel one another, while no poles arise at $z$. For $r=1$, the second factor on the first line has a pole in $x$ at $z$ which is canceled by a zero in $x$ at $z$ in the combination inside the parentheses on the first line.  Similarly, the residues of the poles in $y$ at $z$, arising from the second term inside the parentheses on the first line, and the $k=r$ term on the second line cancel one another. Thus, $S^{I_1 \cdots I_r}{}_K(x,y,z) $ is holomorphic in $x,y,z$ for $r \geq 0$. 

\sm

\textit{(b)} The $\mA$ monodromies vanish because  the $\mA$ monodromies in $x,y$ of the constituents $g^{I_1 \cdots I_r}{}_J(x,y)$ do.

\sm

\textit{(c)} The $\mB$ monodromy (\ref{deltnot}) in $x$ is given as follows, 
\bea 
\Delta _L ^{(x)} S^{I_1 \cdots I_r}{}_K(x,y,z) = A+B+C+D+E
\eea
where we have defined the following shorthands, 
\begin{align}
A & =   \big (  g^{ J}{}_K  (y,x)  - g^{J}{}_K(y,z)   \big ) \Delta _L ^{(x)} g^{I_1 \cdots I_r } {}_J  (x,z) 
\no \\ 
B & =  \Delta _L ^{(x)}  g^{ J}{}_K  (y,x) \big ( g^{I_1 \cdots I_r } {}_K  (x,z)  {+} \Delta _L ^{(x)} g^{I_1 \cdots I_r } {}_K  (x,z)  \big ) 
\no \\ 
C & =   \sum_{k = 1}^r \big ( \Delta _L ^{(x)} g^{  I_1 \cdots I_k } {}_J(x,y) \big )  g^{J I_{k + 1} \cdots I_r}{}_K(y,z)
\no \\ 
D & =  \om_J(y)   \Delta _L ^{(x)} g^{ I_1 \cdots I_r J} {}_K(x,y) 
\no \\ 
E & =  \om_J(y)   \Delta _L ^{(x)} g^{(J \shuffle I_1 \cdots I_{r-1})  I_r} {}_K(x,z) 
\end{align}
and we have used the fact that the $k=0$ contribution in the sum on the third line vanishes, as well as the following expansion of the shuffle product, 
\beq
\label{C.ACD.2}
g^{(J \shuffle I_{\ell+1} \cdots I_{r-1})I_r} {}_K(x,z)
= \sum_{k=\ell}^{r-1} g^{ I_{\ell+1} \cdots I_k J I_{k+1} \cdots I_r} {}_K(x,z) 
\eeq
In $A$, we substitute the expressions (\ref{2.mongx}) for the $\mB$ monodromies in $x$ and recast the result in terms of $S^{I_{\ell+1} \cdots I_r}{}_K(x,y,z)$ plus compensating terms as follows,
\begin{align}
\label{C.A.1}
&A   = 
\sum_{\ell=1}^r { ( - 2 \pi i )^\ell \over \ell!} \delta ^{I_1 \cdots I_\ell} _L S^{I_{\ell+1} \cdots I_r} {}_K(x,y,z)
\no \\ & \ 
- \! \! \sum_{1\leq \ell \leq k}^r \! \! 
 { ( - 2 \pi i )^\ell \over \ell!} \delta ^{I_1 \cdots I_\ell} _L   g^{I_{\ell+1} \cdots I_k}{}_J (x,y) g^{J I_{k+1} \cdots I_r} {}_K(y,z)
\no \\ & \ 
- \om_J(y)  \sum_{\ell=1}^r { ( - 2 \pi i )^\ell \over \ell!} \delta ^{I_1 \cdots I_\ell} _L 
\Big [ g^{I_{\ell+1} \cdots I_r J}{}_K(x,y) \notag \\
&\quad \quad\quad\quad\quad\quad\quad\quad\quad + g^{(J \shuffle I_{\ell+1} \cdots I_{r-1})I_r} {}_K(x,z) \Big ]
\end{align}
Substituting the expressions for the monodromies in  $B$, $C$ and $D$, we obtain, 
\begin{align}
\label{C.C.1}
B & =  - \om_L(y) \sum _{\ell=0}^r { (-2 \pi i )^{\ell+1} \over \ell !} \delta ^{I_1 \cdots I_\ell} _L 
g^{I_{\ell+1} \cdots I_r} {}_K(x,z)
\no \\
C & =   \sum_{1 \leq \ell \leq k}^r
 { (-2 \pi i)^\ell \over \ell!} \delta ^{I_1 \cdots I_\ell}_L g^{I_{\ell+1} \cdots I_k}{}_J(x,y) g^{J I_{k+1} \cdots I_r}{}_K(y,z)
\no \\
D & =  \om_J(y) \sum_{\ell=1}^r { ( - 2 \pi i )^\ell \over \ell!} \delta ^{I_1 \cdots I_\ell} _L 
g^{I_{\ell+1} \cdots I_r J} {}_K(x,y) \notag \\
&\quad
+ \om_L(y) { ( - 2 \pi i )^{r+1} \over (r+1)!} \delta ^{I_1 \cdots I_r }_L \om_K(x)
\end{align}
while $E$ is given by,
\begin{align}
E & =  \om_J(y) \! \sum _{1\leq \ell \leq k}^{r-1} \!
 { ( - 2 \pi i)^\ell \over \ell!} \, \delta ^{I_1 \cdots I_\ell} _L \, g^{I_{\ell+1} \cdots I_k J I_{k+1} \cdots I_r}{}_K(x,z)
\no \\ &\quad
+ \om_L(y) \sum_{\ell=0}^{r-1} { ( - 2 \pi i)^{\ell+1} \over (\ell+1) !} \, \delta ^{I_1 \cdots I_\ell } _L \, g^{I_{\ell+1} \cdots I_r}{}_K(x,z)
\label{sepEeq} \\ &\quad
+ \om_L(y) \sum_{\ell=1}^r  \ell \, 
{ ( - 2 \pi i)^{\ell+1} \over (\ell+1) !} \, \delta ^{I_1 \cdots I_\ell } _L \,  g^{I_{\ell+1} \cdots I_r}{}_K(x,z)
\no
\end{align} 
where the factor of $\ell$ in the summand of the last line of $E$ originates from the $\ell$ equal contributions that arise from the shuffle product $\delta ^{(J \shuffle I_1 \cdots I_{\ell-1} ) I_\ell }_L$.  

\sm

To prove the first equation in (\ref{B.lem.1})  we show that in the sum of $A$, $B$, $C$, $D$, and $E$ all terms other than the first line in (\ref{C.A.1}) cancel one another. We see that $C$ cancels the second term in $A$, while the first term in $D$ cancels the first term inside the brackets on the third line of $A$. Using the fact that the $\ell=r$ contribution to the second term inside the brackets of the last line in (\ref{C.A.1}) vanishes, we see that the shuffle products cancel the double sum over $k$ and $\ell$ in the first line of $E$ in (\ref{sepEeq}). Moreover, the contributions for $\ell \leq r-1$ from $B$ and the last two lines of $E$ cancel one another. Finally, the $\ell=r$ contributions of $B$ and the last line of $E$ cancel the contribution from the second term in $D$, thereby completing the proof of~(\ref{B.lem.1}). 

\sm

Proving  the vanishing of the $\mB$ monodromies in $y$ and the formula for the $\mB$ monodromies in $z$ in (\ref{B.lem.1}) proceeds along analogous lines and is left to the reader. This concludes the proof of the Lemma.  \end{proof}

\vspace{\mylength}
\subsection{Proof of the Fay identity using the Lemma}
\vspace{\myotherlength}

The proof of the Fay identity (\ref{13.a}), which states that $S^{I_1 \cdots I_r}{}_K(x,y,z)=0$ for all $r \geq 1$,  
proceeds along the same line of arguments used to prove Theorem 9.2 of~\cite{DHoker:2024ozn}, which established the generalized meromorphic interchange lemma. Here as well, the key ingredients are the holomorphicity and the monodromies of $S^{I_1 \cdots I_r}{}_K(x,y,z)$ in the variable $x$ established in the above Lemma. We cut the Riemann surface $\Sigma$ along a set of canonical homology cycles that share a common base point $q \in \Sigma$ as in figure \ref{fig:1}, and decompose the boundary $\mC = \partial D$ of the resulting fundamental domain $D$ as follows,
\bea
\mC = \bigcup _{L=1}^h \mA^L \star \mB_L \star (\mA^L)^{-1} \star (\mB_L)^{-1} 
\eea
where $\mC_1 \star \mC_2$ denotes the composed path that traverses first $\mC_1$ followed by $\mC_2$. We then use the holomorphicity in $x$ of $S^{I_1 \cdots I_r}{}_K(x,y,z)$ to conclude that its Cauchy integral around $\mC$ must vanish,
\bea
\oint _\mC S^{I_1 \cdots I_r}{}_K(t,y,z) =0
\eea
The contributions from the integrals over $\mB_L$ and $(\mB_L)^{-1}$ cancel one another as these curves are related to one another by the $\mA^K$ monodromy transformation that maps $\mB_L$ to $(\mB_L)^{-1}$ while reversing its orientation. The remaining contributions take the form, 
\beq
\sum_{L=1}^h \oint _{\mA^L} \Big ( S^{I_1 \cdots I_r}{}_K (\mB_L \cdot t,y,z) - S^{I_1 \cdots I_r}{}_K(t,y,z) \Big ) =0
\eeq
But this monodromy is given by (\ref{B.lem.1}) and we get, 
\beq
\label{B.4}
\sum_{\ell=1}^{r-1} { (-2 \pi i )^\ell \over \ell !} \sum_{L=1}^h 
\delta ^{I_1 \cdots I_\ell}_L \oint _{\mA^L} S^{I_{\ell+1} \dots I_r} {} _K(t,y,z)  =0
\eeq 
where we have used $S^\emptyset {} _K(x,y,z)=0$ to omit the vanishing $\ell=r$ contribution. Since $S^\emptyset {} _K(x,y,z)=0$, equation (\ref{B.lem.1}) implies that $S^{I_1}{}_K(x,y,z)$ has vanishing $\mA$ and $\mB$ monodromies in $x$ and is thus a single-valued $(1,0)$ form in $x$. Specializing  to the case $r=2$ reduces (\ref{B.4})  to a single term from $\ell=1$ given by, 
\bea
\oint _{\mA^{I_1} }  S^{I_2} {} _J(t,y,z)  =0
\eea
Since $S^{I_2} {} _J(x,y,z) $ is a single-valued holomorphic $(1,0)$ form with vanishing $\mA$-periods, it must vanish. 
We now proceed by induction on $r$ and assume that $S^{I_1 \cdots I_k}{}_K(x,y,z)=0$ for all $k \leq r-1$. The above Lemma then implies that $S^{I_1 \cdots I_r}{}_K(x,y,z)$ is a single-valued holomorphic  $(1,0)$ form in $x$. Substituting $r \to r+1$ then reduces (\ref{B.4}) to a single term $\ell=1$,
\bea
 \oint _{\mA^{I_1} } S^{I_{2} \dots I_{r+1}} {} _K(t,y,z)  =0
\eea
Since the indices $I_1 \cdots I_{r+1}$ are arbitrary, the above relation implies $S^{I_1 \cdots I_r}{}_K(x,y,z)=0$, thus completing the proof by induction of (\ref{13.a}).

\vspace{\mylength}
%%%%%%%%%%%%%%%%%%%%%%%%%%%%%%%%%%%%%%%%%%%
%%%%%%%%%%%%%%%%%%%%%%%%%%%%%%%%%%%%%%%%%%%
\section{Appendix B: The effects of poles crossing $\mA$ cycles }
\label{sec:B}
%%%%%%%%%%%%%%%%%%%%%%%%%%%%%%%%%%%%%%%%%%%
%%%%%%%%%%%%%%%%%%%%%%%%%%%%%%%%%%%%%%%%%%%
\vspace{\myotherlength}

Theorem 2 provides a recursion relation for the Enriquez kernels $g^{I_1 \cdots I_r}{}_J(x,y)$ in terms of integrals over $\mA$ cycles of lower rank kernels. As soon as $r \geq 2$, this relation involves  integrals over multiple $\mA$ cycles, which were defined in (\ref{limit}). The two-fold goal of this appendix is to 
\begin{itemize}
\item[(i)] illustrate the necessity of displacing $\mA^{L}_\ep $ in (\ref{limit}) into the interior of the fundamental domain, and to 
\item [(ii)] verify
the new representation (\ref{4.thm.2}) of Enriquez kernels to reproduce the required $\mB$ monodromies~(\ref{2.mongx}).
\end{itemize} 
Both of (i) and (ii) will be accomplished by evaluating contributions that are caused by a pole crossing an $\mA$ cycle: (i) the effect of swapping the order of integration over $\mA$ cycles, i.e.\ departing from the
limit prescribed in (\ref{limit}); (ii) the consistent matching of monodromies on both sides of~(\ref{4.thm.2}).

\sm

(i) Swapping the order of integration involves  
subtle corrections that we will now illustrate for a 
double integral of the same type that occurs in (\ref{limit}), 
\bea
\label{B.9}
 \lim _{\ep \to 0} \oint ^{(t)} _{\mA^{L}_\ep }  \oint ^{(t')} _{\mA^{M}} 
g^J{}_K(t',t)  g^I{}_J(t,z)  & = & - \pi^2 \delta^{ILM}_K
\no \\ 
 \lim _{\ep \to 0} \oint ^{(t')} _{\mA^{M}_\ep }  \oint ^{(t)} _{\mA^{L}} 
g^J{}_K(t',t)  g^I{}_J(t,z) & = & + \pi^2 \delta ^{ILM}_K
\qquad
\eea
Here and below, the superscript of $ \oint ^{(t)} _{\mA^{L} } $ specifies the variable $t$
which is integrated over the cycle $\mA^{L}$ in the subscript.
In the first line of (\ref{B.9}) the integral over $t'$, carried out using (\ref{1.per.g}), is independent of $t$ so that the integral over $t$ may be performed using (\ref{1.per.g}). In the second line of (\ref{B.9}) we appeal to Theorem 2 for the integration over $t$ after which the integral over $t'$ may be carried out using (\ref{1.per.g}). Since the contribution is non-vanishing only for $M=L$, the difference between the first and second integral in (\ref{B.9}) may be explained by a residue calculation,
\bea
\oint _{\mA^L}^{(t)} \Big ( 2 \pi i \, \mathop{\mathrm{Res}} _{ t'=t} g^J{}_K(t',t) \Big ) g^I{}_J(t,z)
 = - 2 \pi^2 \delta ^{IL}_K
\eea

(ii) To verify that the monodromies in $y$ around a cycle $\mB_M$ of both sides of (\ref{4.thm.2}) match, we use the fact that the term on the left of (\ref{4.thm.2}) and the second line  on its right side transform according to  (\ref{2.mongx}). The monodromy of the integral in (\ref{4.thm.2}) is given by a combination of the monodromy  of $g^J{}_K(y,t)$ and the effect that, upon moving $ y \to \mB_M \cdot y$, the pole in $t$ at $y$ produces a residue as it crosses the branch cut $\mA^L$ provided that these two cycles intersect, i.e.\ $M=L$. We arrive at (see (\ref{deltnot}) for the $\Delta ^{(y)}_M$ notation),  
\begin{align}
\Delta ^{(y)}_M  \oint _{\mA^L} & g^{ J}{}_K  (y,t) \, g^{I_1 \cdots I_r } {}_J  (t,z)
\no \\
& =  (- 2 \pi i)^{r+1} {B_r \over r!}  \,  \delta ^{I_1 \cdots I_r L}_M \, \om_K  (y) 
\no \\
& \quad - 2 \pi i \, \delta ^L_M \, g^{I_1 \cdots I_r}{}_K (y,z)
\end{align}
Combining all contributions, the monodromies are indeed found to match. Developing the corresponding argument for the monodromies in $z$ and verifying agreement with the $\mB$ monodromies in (\ref{2.mongy}) is left to the reader.

\vspace{\mylength}
%%%%%%%%%%%%%%%%%%%%%%%%%%%%%%%%%%%%%%%%%%%
%%%%%%%%%%%%%%%%%%%%%%%%%%%%%%%%%%%%%%%%%%%
\section{Appendix C: Proof of Theorem 4}
\label{sec:C}
%%%%%%%%%%%%%%%%%%%%%%%%%%%%%%%%%%%%%%%%%%%
%%%%%%%%%%%%%%%%%%%%%%%%%%%%%%%%%%%%%%%%%%%
\vspace{\myotherlength}

In this section we prove Theorem 4 and the variational formula (\ref{5.thm.1}) for arbitrary $r \geq 1$. To do so, we begin by obtaining a recursion relation for $\delta _{ww} g^{I_1 \cdots I_r}{}_J(x,y)$ by  taking the $\delta_{ww}$ variation of the recursion relation (\ref{4.thm.2}) with the help of the $r=1$ case (\ref{14.fin.2}) and the variational formulae in (\ref{14.a}), 
\bea
\label{C.1}
0 & = & \oint _{\mA^L}  \Big ( \p_w \p_x \ln E(w,x)  \, g^J{}_K(w,t) 
 \\ && \hskip 0.4in
+  \om_M(w) \, \p_w g^{JM}{}_K (x,w) \Big ) g^{I_1 \cdots I_r } {}_J  (t,y)  
\no \\ &&
+ \oint _{\mA^L}  g^{ J}{}_K  (x,t)  \,  \delta_{ww}  g^{I_1 \cdots I_r } {}_J  (t,y) 
\no \\ &&
+ \sum_{k = 0}^{r-1} (-2 \pi i)^k { B_k \over k!} \, \delta ^{I_1 \cdots I_k }_L \,  \delta_{ww}  g^{L I_{k + 1} \cdots I_r}{}_K(x,y) 
\no \\ &&
+  \om_K(w) \p_w \p_x \ln E(w,x)   (-2 \pi i)^{r+1} { B_{r+1} \over r!} \, \delta ^{I_1 \cdots I_r  L}_K
\no
\eea
where the $k=0$ term of the fourth line accounts for the $\delta_{ww}$ variation of the left-hand side of (\ref{4.thm.2}). Combining the contribution from the first term inside the parentheses of the first line with the last line we recognize the combination occurring in the first and third terms of (\ref{4.thm.2}) with 
$y$ replaced by $w$ and $z$ replaced by $y$. By also carrying out the integral over $t$ of the terms on the second line we~obtain,  
\bea
\label{C.2}
0 & = & 
 \oint _{\mA^L}  g^{ J}{}_K  (x,t)  \Big ( \delta_{ww}  g^{I_1 \cdots I_r } {}_J  (t,y)   \Big )
 \\ &&
 + \om_M(w) \, \p_w g^{LM}{}_K (x,w)  (- 2 \pi i)^r {B_r \over r!} \delta ^{I_1 \cdots I_r }_L 
\no \\ &&
+ \sum_{k = 0}^{r-1} (-2 \pi i)^k { B_k \over k!} \, \delta ^{I_1 \cdots I_k }_L 
\Del_{ww} \, g^{L I_{k + 1} \cdots I_r}{}_K(x,y) 
\no
\eea
where $\Del_{ww}$ is defined as follows,
\bea
\Del_{ww} g^{I_1 \cdots I_r } {}_J  (x,y) & = & \delta _{ww} g^{I_1 \cdots I_r } {}_J  (x,y)
 \\ &&
- \p_w\p_x \ln E(w,x) g^{I_1 \cdots I_r } {}_J  (w,y)
\no 
\eea
Expressing the integrand in the first term of (\ref{C.2}) in terms of the quantity $\Del_{ww} \, g^{I_1 \cdots I_r}{}_J(x,z) $ we obtain,
\bea
\label{C.3}
0 & = & 
 \oint  _{\mA^L}  g^{ J}{}_K  (x,t)  \Del_{ww} \, g^{I_1 \cdots I_r } {}_J  (t,y)  
 \\ &&
+\, g^{I_1 \cdots I_r}{}_J(w,y)  \oint _{\mA^L}  g^{ J}{}_K  (x,t)  \p_w \p_t \ln E(w,t) 
\no \\ &&
+\, \sum_{k = 0}^{r-1} (-2 \pi i)^k { B_k \over k!} \, \delta ^{I_1 \cdots I_k }_L \Del_{ww} \,  g^{L I_{k + 1} \cdots I_r}{}_K(x,y)  
 \no \\ &&
+\,  \om_M(w) \, \p_w g^{JM}{}_K (x,w)  (- 2 \pi i)^r {B_r \over r!} \delta ^{I_1 \cdots I_r L}_J 
\no
\eea
The integral on the second line evaluates as follows,
\bea
\oint _{\mA^L}  g^{ J}{}_K  (x,t)  \p_w \p_t \ln E(w,t) 
= - \p_w g^{LJ}{}_K(x,w)
\eea
as one can check be casting the left side into the form
$\partial_w \oint _{\mA^L} g^I{}_K(x,t) g^J{}_I(t,w)$ via (\ref{3.thm.1}) 
and using (\ref{4.thm.2}) at $r=1$.
 In summary we obtain the recursion relation, 
\begin{align}
\label{C.4}
 \oint  _{\mA^L} & g^{ J}{}_K  (x,t)  \Del_{ww} \, g^{I_1 \cdots I_r } {}_J  (t,y)  
 \\ &
+ \sum_{k = 0}^{r-1} (-2 \pi i)^k { B_k \over k!} \, \delta ^{I_1 \cdots I_k }_L \Del_{ww} \,  g^{L I_{k + 1} \cdots I_r}{}_K(x,y)  
\no \\ = \, &
 \p_w g^{LJ}{}_K(x,w)  g^{I_1 \cdots I_r}{}_J(w,y) 
 \no \\ &
-  \om_M(w) \, \p_w g^{JM}{}_K (x,w)  (- 2 \pi i)^r {B_r \over r!} \delta ^{I_1 \cdots I_r L}_J 
\no 
\end{align}
Iterating the recursion relation (\ref{C.4}) to the lowest few orders,  we obtain, 
\bea
\label{14.Dww.1}
\Del_{ww} \, g^I{}_J(x,y) & = & \p_w g^{IM}{}_J(x,w) \om_M(w)
 \\
\Del_{ww} \, g^{LI}{}_J(x,y) & = &  \p_w  g^{LIM}{}_J(x,w) \om_M(w) 
\no \\ &&
+  \p_w g^{LM}{}_J(x,w)  g^{I}{}_M(w,y) 
\no
\eea
These results agree with the general formula in (\ref{5.thm.1}). 

\sm

Theorem 4 for arbitrary $r\geq 1$ is equivalent to the vanishing of  the following function,
\begin{align}
\tilde \Del_{ww}  \, & g^{I_1 \cdots I_r}{}_J (x,y)  = \Del_{ww} \, g^{I_1 \cdots I_r}{}_J(x,y)  
\label{eqthm4} \\ &
- \sum_{k=1}^r \p_w g^{I_1 \cdots I_k M}{}_J(x,w) \, g^{I_{k+1} \cdots I_r}{}_M(w,y)
\no 
\end{align}
for all $r \geq 1$. We proceed by re-expressing both instances of $\Del_{ww} \, g$ in (\ref{C.4}) in terms of $\tilde \Del_{ww} \, g$. Reformulating the result in terms of the following combination,
\bea
\cA_r & = & 
\oint  _{\mA^L}  g^{ J}{}_K  (x,t)  \tilde \Del_{ww} \, g^{I_1 \cdots I_r } {}_J  (t,y)  
 \\ &&
+ \sum_{k = 0}^{r-1} (-2 \pi i)^k { B_k \over k!} \, \delta ^{I_1 \cdots I_k }_L 
\tilde \Del_{ww} \,  g^{L I_{k + 1} \cdots I_r}{}_K(x,y) 
\no
\eea
we obtain, 
\bea
\cA_r & = & 
- \sum_{\ell=1}^r \oint_{\mA^L} g^J{}_K(x,t) \p_w g^{I_1 \cdots I_\ell M}{}_J(t,w) 
\label{carexp} \\ && \hskip 1.5in 
\times g^{I_{\ell+1} \cdots I_r}{}_M(w,y)
\no \\ &&
- \sum_{k=0}^{r} \sum _{\ell=k}^r (-2 \pi i)^k {B_k \over k!} \delta ^{I_1 \cdots I_k}_L 
 \p_w g^{L I_{k+1} \cdots  I_\ell M}{}_K(x,w) 
 \no \\ && \hskip 1.5in 
\times g^{I_{\ell+1} \cdots I_r}{}_M(w,y)
\no \\ && 
+ \p_w g^{LJ}{}_K(x,w) g^{I_1 \cdots I_r}{}_J(w,y)
\no
\eea
The integral on the first line evaluates as follows,
\begin{align}
\oint _{\mA^L} & g^J{}_K(x,t)  \p_w g^{I_1 \cdots I_\ell M}{}_K(t,w) 
 \\ &= 
- \sum_{k=0}^\ell (-2 \pi i )^k {B_k \over k!} \delta ^{I_1 \cdots I_k} _L \p_w g^{L I_{k+1} \cdots I_\ell M}{}_K(x,w)
\no
\end{align}
Substituting this result into $\cA_r$~gives, 
\bea
\cA_r & = & 
 \sum_{\ell=1}^r \sum_{k=0}^\ell (-2 \pi i )^k {B_k \over k!} \delta ^{I_1 \cdots I_k} _L 
 \\ && \hskip 0.5in
\times \p_w g^{L I_{k+1} \cdots I_\ell M}{}_K(x,w) g^{I_{\ell+1} \cdots I_r}{}_M(w,y)
\no \\ &&
- \sum_{k=0}^{r} \sum _{\ell=k}^r (-2 \pi i)^k {B_k \over k!} \delta ^{I_1 \cdots I_k}_L 
\no \\ && \hskip 0.5in 
\times \p_w g^{L I_{k+1} \cdots  I_\ell M}{}_K(x,w) g^{I_{\ell+1} \cdots I_r}{}_M(w,y)
\no \\ && 
+ \p_w g^{LJ}{}_K(x,w) g^{I_1 \cdots I_r}{}_J(w,y)
\no
\eea
Interchanging the summations in the second double sum according  to 
$\sum_{k=0}^{r} \sum _{\ell=k}^r = \sum _{\ell=0}^r \sum_{k=0}^\ell$, 
we see that the $\ell\geq 1$ terms of the third and fourth line cancel the first two lines, while the $\ell=0$ contribution on the third and fourth line cancels the last line. Thus, we find, $\cA_r = 0$ for all $r \geq 1$, or equivalently,
 \bea
0 & = & \oint  _{\mA^L}  g^{ J}{}_K  (x,t)  \tilde \Del _{ww} \, g^{I_1 \cdots I_r } {}_J  (t,y)  
\label{ointar} \\ &&
+ \sum_{k = 0}^{r-1} (-2 \pi i)^k { B_k \over k!} \, \delta ^{I_1 \cdots I_k }_L \tilde \Del_{ww} \,  g^{L I_{k + 1} \cdots I_r}{}_K(x,y) 
\no
\eea
Isolating the $k=0$ term, we may re-express this equation as a recursion relation,
\begin{align}
\! \! \! \tilde \Del_{ww} & \, g^{L I_1 \cdots I_r } {}_J  (x,y)   = -  \oint  _{\mA^L}  g^{ J}{}_K  (x,t)  \tilde \Del_{ww} \, g^{I_1 \cdots I_r } {}_J  (t,y)  
\no \\
& \! \! \! -  \sum_{k = 1}^{r-1} (-2 \pi i)^k { B_k \over k!} \, \delta ^{I_1 \cdots I_k }_L \tilde \Del_{ww} \,  g^{L I_{k + 1} \cdots I_r}{}_K(x,y) 
\end{align}
The number of upper indices of $\tilde \Del_{ww} g$ on the left side is $r+1$ and therefore strictly
larger than the at most $r$ upper indices of the combinations $\tilde \Del_{ww} g$ on the right side.  Having already proven in (\ref{14.c}) to (\ref{14.fin.2}) that $\tilde \Del_{ww} g^{I_1 \cdots I_r}{}_J(x,y)$ vanishes for $r=1$, it follows that the combination vanishes for all $r$, thus completing the proof of Theorem 4.

\vspace{\myotherlength}

%{99}

\end{document}